\documentclass[onecolumn,amsmath,amssymb,11pt,superscriptaddress,nofootinbib]{revtex4-2}
\usepackage[T1]{fontenc}
\usepackage{lmodern}
\usepackage[british]{babel}
\makeatletter\AtBeginDocument{\let\@elt\relax}\makeatother
\usepackage{amsmath}
\usepackage{amssymb}
\usepackage{physics}
\usepackage{amsthm}
\usepackage[]{graphicx}
\usepackage[justification=centering]{caption}
\usepackage{subcaption}
\usepackage{tensor}
\usepackage[dvipsnames]{xcolor}
\usepackage{cancel}
\usepackage{setspace}
\usepackage{fancyhdr}
\usepackage{appendix}
\usepackage{url}
\usepackage{natbib}
\usepackage{hyperref}
\hypersetup{
	colorlinks = true,
	linkbordercolor = {white},
	linkcolor=black,
	citecolor=blue,
	urlcolor=black}

\def\be{\begin{equation}}
\def\ee{\end{equation}}
\def\beq{\begin{eqnarray}}
\def\eeq{\end{eqnarray}}
\newcommand{\pkt}{\; .}
\newcommand{\kma}{\; ,}

\def\e{{\rm e}}

\begin{document}

\allowdisplaybreaks

\title{A Zoo of Axionic Wormholes \vspace{0.5cm}}

\author{Caroline Jonas}
\email[]{caroline.jonas@aei.mpg.de}
\affiliation{Max Planck Institute for Gravitational Physics \\ (Albert Einstein Institute), 14476 Potsdam, Germany}
\author{George Lavrelashvili}
\email[]{george.lavrelashvili@tsu.ge}
\affiliation{Department of Theoretical Physics, A.Razmadze Mathematical Institute \\
	at I.Javakhishvili Tbilisi State University, GE-0193 Tbilisi, Georgia}
\author{Jean-Luc Lehners}
\email[]{jlehners@aei.mpg.de}
\affiliation{Max Planck Institute for Gravitational Physics \\ (Albert Einstein Institute), 14476 Potsdam, Germany}

\begin{abstract}
\vspace{1cm}
As was discovered some time ago by Giddings and Strominger (GS), an axion can support a wormhole geometry in the presence of a massless dilaton, as long as the dilaton coupling remains below a critical value. We find that when the dilaton becomes massive, the set of solutions is vastly increased: not only do solutions exist above the critical value of the coupling, but new branches of solutions with several minima in the geometry also appear. All of these generalised GS-like solutions possess the property that, when analytically continued, they lead to a contracting baby universe. We show that in addition there exist families of solutions which, upon analytic continuation, lead to expanding baby universes. A curious property of axion-dilaton wormhole families is that their Euclidean action often decreases when the solutions acquire additional oscillations in the fields. When we replace the dilaton by an ordinary scalar field with a double well potential, we find analogous wormhole families leading to expanding baby universes. This time the Euclidean action has the expected behaviour of increasing with the number of oscillations in the fields, although it also contains a puzzling aspect in that some solutions possess a negative action.
\end{abstract}

\maketitle

\newpage

\tableofcontents


\section{Introduction}

The questions of the possibility and consequences of topology change have haunted quantum gravity research for several decades now. The simple fact that in a theory of quantum gravity the spacetime manifold should be able to fluctuate suggests that the overall topology of the manifold might also vary and that, on the smallest scales, one might have to think of spacetime as having a foam-like structure \cite{Wheeler:1957mu,Carlip:2022pyh}. With the discovery of explicit wormhole solutions sourced by axions \cite{Giddings:1987cg}, a number of questions became more concrete. For instance, if baby universes can be created locally, then they can carry information away, leading to apparent non-unitary behaviour \cite{Hawking:1987mz,Lavrelashvili:1987jg,Lavrelashvili:1988jj,Giddings:1988cx}. Also, if wormholes can connect distant spatial regions, then their presence will influence the value of coupling constants, effectively leading to random values of said ``constants''. Closer inspection of this question led Coleman to the picture of $\alpha-$vacua, describing superselection sectors of quantum gravity within which quantum coherence would be individually retained \cite{Coleman:1988cy}. 

More recently, these questions have received renewed interest, as it was realised that puzzling features remained: for instance, it is a general expectation that quantum gravity should admit no free parameters \cite{Banks:2010zn}, and this expectation is in obvious conflict with the above-mentioned $\alpha-$parameters. Also, in the context of the AdS/CFT correspondence, if there are multiple disconnected boundary regions, then one would expect the CFT partition function to factorise correspondingly. However, if wormholes can connect the asymptotic regions, then they lead to interactions mediated via the bulk -- again we encounter a paradox \cite{Rey:1998yx,Maldacena:2004rf,Arkani-Hamed:2007cpn}. Possible resolutions have been suggested: on the one hand, wormholes might admit negative modes and thus not contribute to the gravitational path integral \cite{Hertog:2018kbz} (though for purely axionic wormholes, the recent work \cite{Loges:2022nuw} casts doubt on this possibility by showing that these particular wormholes do not admit negative modes). On the other hand, it has been suggested that in the gravitational path integral there might be a huge degeneracy arising from the interplay of states of various topologies \cite{Marolf:2020xie}, perhaps effective enough to reduce the baby universe Hilbert space to a single dimension and obviate the need for $\alpha-$parameters \cite{McNamara:2020uza}.

\begin{figure}[h]
	\centering
	\begin{subfigure}{0.49\linewidth}
		\includegraphics[width=6cm]{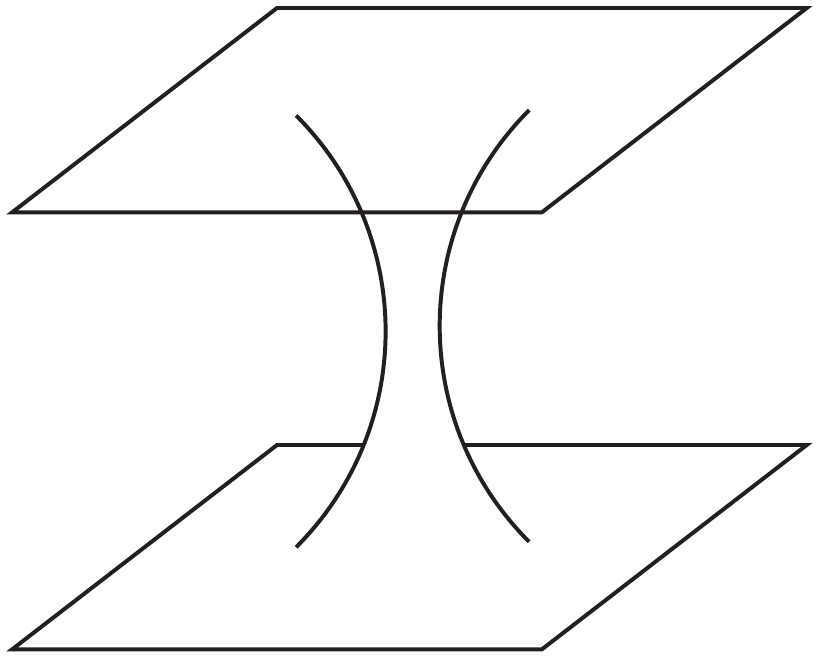}
	\end{subfigure}
	\begin{subfigure}{0.49\linewidth}
		\includegraphics[width=6cm]{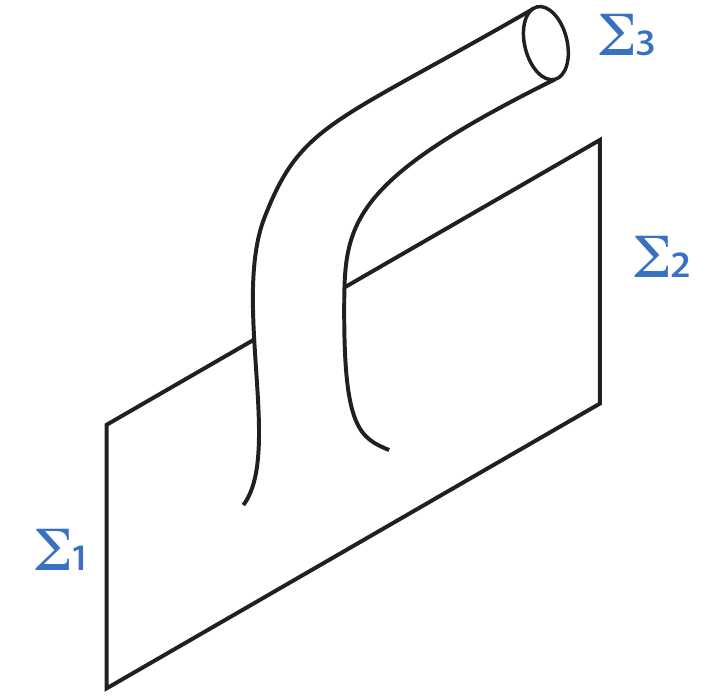}
	\end{subfigure}
	\caption{\small Common interpretations of Euclidean wormholes. On the left, a full wormhole connecting two asymptotic regions. On the right, a semi-wormhole, leading to the creation of a baby universe ($\Sigma_3$).}\label{fig:wormholeinstanton}
\end{figure}

\begin{figure}[h]
	\centering
	\begin{subfigure}{0.31\linewidth}
		\includegraphics[width=\textwidth]{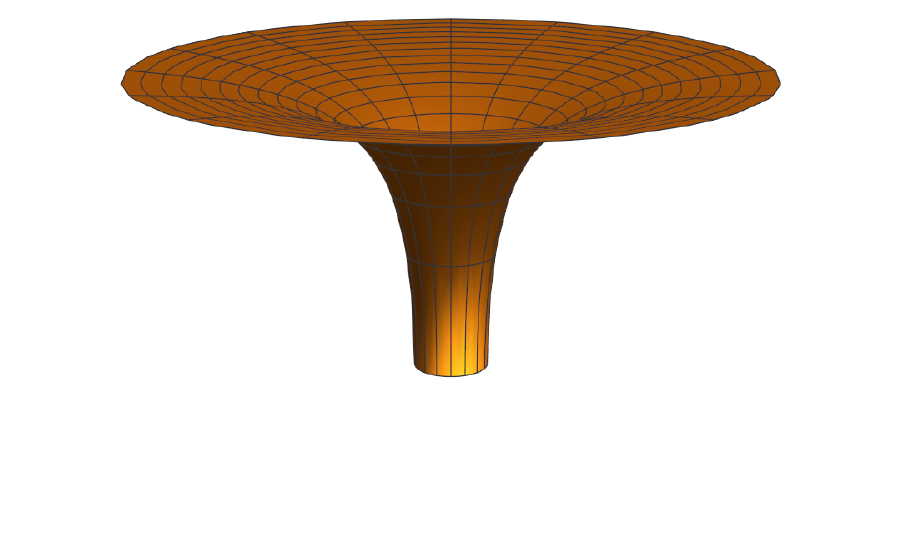}\caption{Giddings-Strominger-type wormhole.}\label{fig:GSwormhole}
	\end{subfigure}
	\begin{subfigure}{0.31\linewidth}
		\includegraphics[width=\textwidth]{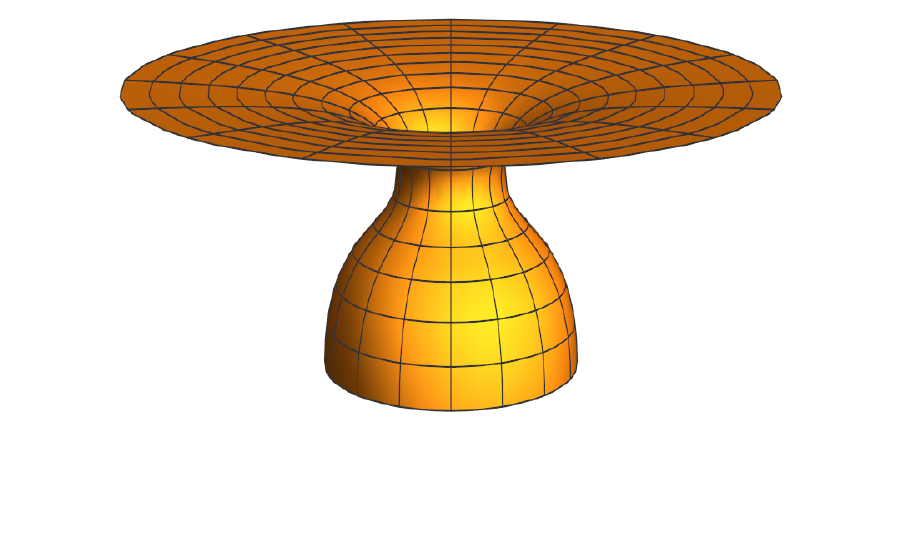}\caption{Wormhole leading to an expanding baby universe.}\label{fig:ourwormhole}
	\end{subfigure}
	\begin{subfigure}{0.31\linewidth}
		\includegraphics[width=\textwidth]{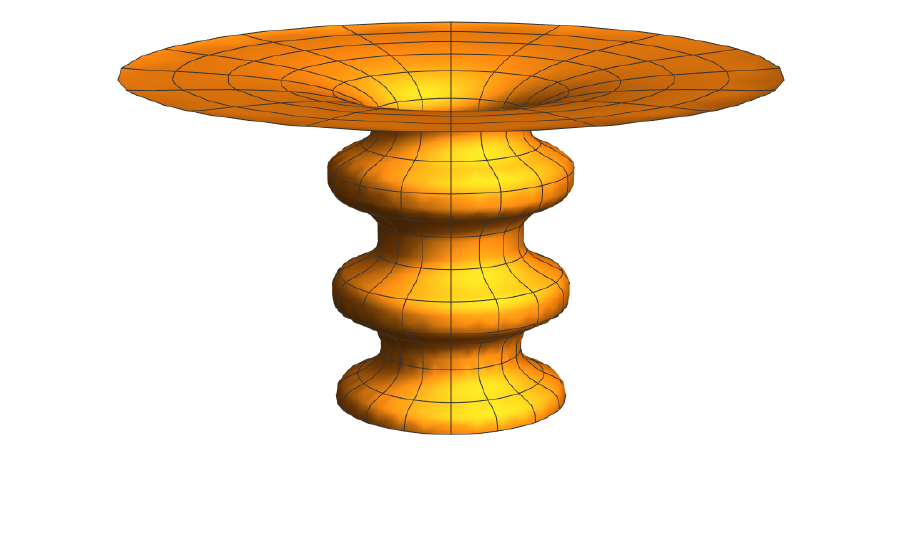}\caption{An oscillating wormhole leading to an expanding baby universe.}\label{fig:oscwormhole}
	\end{subfigure}
	\caption{\small Visualisation of Euclidean wormholes. Wormholes of GS type lead to contracting baby universes upon analytic continuation, while wineglass shaped wormholes lead to expanding baby universes. The right panel shows the oscillating solution with $2$ extra minima of the scale factor included in Fig.~\ref{fig:oscillations}. See Appendix \ref{sec:Embedding diagrams} for further details on embedding diagrams of the sort shown here.}\label{fig:embeddingdiagram}
\end{figure}

These puzzles remain unresolved to date, and motivate us to further study the existence and properties of wormholes (see also \cite{Hebecker:2016dsw,Alonso:2017avz,Marolf:2021kjc,Andriolo:2022rxc} for related recent works, and \cite{Hebecker:2018ofv,Kundu:2021nwp} for useful reviews). A full wormhole solution can be thought of \cite{Giddings:1987cg,Alonso:2017avz,Hebecker:2016dsw}
as a (gravitational) instanton-anti-instanton pair connecting two asymptotic regions, whereas
a semi-wormhole is interpreted as a gravitational instanton mediating a topology change $\Sigma_1 \to \Sigma_2\oplus \Sigma_3$,
{\it e.g.} $\mathbb{R}^3 \to \mathbb{R}^3 \oplus S^3$,
and leading to the creation of a baby universe, see Fig.~\ref{fig:wormholeinstanton}. It was noted in \cite{Lavrelashvili:1988un} that Gidding-Strominger (GS) wormholes (such as those depicted in Fig.~\ref{fig:GSwormhole}) lead to contracting baby universes after analytic continuation to Minkowski time. In contrast, wormholes leading to expanding baby universes
should have a wineglass shaped neck, as that depicted in Fig.~\ref{fig:ourwormhole}.
Such wormholes were first found in \cite{Lavrelashvili:1988un} in a theory of axion-scalar gravity 
with an  asymmetric double well scalar potential, and further elaborated on in \cite{Rubakov:1988wx}.

In the present paper we describe many new families of wormhole solutions, both in axion-dilaton gravity with a massive dilaton (greatly extending the results of \cite{Andriolo:2022rxc}) and in axion-scalar theory with a symmetric double well potential. We classify the solutions with regard to the distinction mentioned above, namely whether  after analytic continuation they lead to contracting baby universes, or to expanding baby universes. Much less is known about the latter class, but even for the former we uncover a surprisingly intricate structure, for a preview see Fig.~\ref{fig:branchstructureGS}. In all cases we calculate the Euclidean action and find some unexpected features, which we will describe in detail below. Upon variation of the parameters of the theory and/or the axion charge, the wormhole solutions can develop oscillations both in the scalar/dilaton field and in the scale factor of the universe, see Fig.~\ref{fig:oscwormhole}. The physical significance of these oscillations remains to be clarified, in particular as it is related to some of the surprising features of the Euclidean action that we find. At critical values of the parameters, entire new families of solutions emerge in bifurcating patterns, a property which for instance implies that these new branches of solutions are not continuously connected to the massless limit of the axion-dilaton theory and thus explains why these wormhole solutions have gone unnoticed so far. Overall, our main realisation is that significantly more wormhole solutions exist than hitherto suspected. It is our hope that a better understanding of these concrete wormhole solutions will help in elucidating the conceptual puzzles described above.


\section{Axion-dilaton and axion-scalar gravitational theories}

\subsection{Action, field equations and ansatz}

Our starting point is the Euclidean action for gravity coupled to an axion and a dilaton/scalar $\phi$, which reads \cite{Andriolo:2022rxc}:
\begin{equation}
S_\text{E}=\int\dd^4x\sqrt{g}\left(-\frac{1}{2\kappa}R+\frac{1}{2}\nabla_{\mu}\phi \nabla^{\mu}\phi
+V(\phi)+ \frac{1}{12 f^2} e^{-\beta \phi\sqrt{\kappa}} H_{\mu\nu\rho}H^{\mu\nu\rho} \right)
\,,
\end{equation}
where $\kappa\equiv M_\text{Pl}^{-2}=8\pi G$, the dilatonic coupling constant is denoted $\beta$ and the potential $V(\phi)$, $H_{\mu\nu\rho}$ being the 3-form field strength of an axion field with coupling $f$. When $\beta \neq 0,$ we refer to $\phi$ as a dilaton, while for $\beta=0$ we simply call it a scalar. Note that we do not add a Gibbons-Hawking-York boundary term to the action, as we will discuss in more detail below. The corresponding equations of motions are
\begin{equation}
	\left\lbrace
	\begin{aligned}
		&R_{\mu\nu}=\kappa \partial_\mu \phi\partial_\nu \phi + \kappa V g_{\mu\nu} +\frac{\kappa}{2 f^2} \e^{-\beta \phi\sqrt{\kappa}} H_{\mu\rho\sigma} H_\nu^{\rho\sigma}-\frac{\kappa}{6 f^2}\e^{-\beta \phi\sqrt{\kappa}} H_{\gamma\rho\sigma}H^{\gamma\rho\sigma} g_{\mu\nu} \kma \\
		&\frac{1}{\sqrt{g}}\partial_\mu(\sqrt{g}g^{\mu\nu} \partial_\nu \phi)= \frac{\partial V}{\partial\phi}
		- \frac{\beta\sqrt{\kappa}}{12 f^2} 		\e^{-\beta \phi\sqrt{\kappa}} H_{\gamma\rho\sigma}H^{\gamma\rho\sigma}\kma\\
		&\partial_\mu (\sqrt{g} \e^{-\beta\phi\sqrt{\kappa}} H^{\mu\rho\sigma})= 0 \pkt
	\end{aligned}
	\right.\label{eq:eomfullaxiongravitydilaton}
\end{equation}

We will focus on the following spherically symmetric and homogeneous ansatz
\begin{equation}
	\left\lbrace
	\begin{aligned}
	&\dd s^2=h^2(\tau) \dd \tau^2 + a(\tau)^2 \dd\Omega_3^2\,,\\
	&\phi=\phi(\tau)\,,\\
	&H_{0ij}=0\,,\ H_{ijk}=q \varepsilon_{ijk}
\kma
	\end{aligned}
	\right.\label{eq:sphericalansatz}
\end{equation}
which leads to the reduced action
\beq
S_\text{E}^\text{red}
&=&2 \pi^2 \int\dd\tau \left(-\frac{3a\dot{a}^2}{\kappa h}
+ \frac{a^3 \dot{\phi}^2}{2 h}-\frac{3 a h}{\kappa} +h a^3 V + \frac{N^2 h}{a^3}e^{-\beta \phi\sqrt{\kappa}}
\right) + 2 \pi^2 \int\dd\tau \frac{\dd}{\dd\tau} \left(\frac{3a^2\dot{a}}{\kappa h}\right),\label{eq:redact}
\eeq
where
\be
\displaystyle N^2 \equiv \frac{q^2}{2 f^2} \kma
\ee
and the overdot denotes a derivative with respect to the (Euclidean) time coordinate $\tau$. Note that the axion charge $q$ is quantised in string theory (when seen as a source for branes, this follows from the associated Dirac quantisation condition \cite{Henneaux:1986ht}), and hence $q$ and consequently also $N$ should be thought of as being proportional to integers. The surface term arose from integration by parts, and is important in order to obtain the correct value of the action.
Varying the reduced action \eqref{eq:redact} with respect to $a,\,h$ and $\phi$ yields the following equations of motion (strictly equivalent to \eqref{eq:eomfullaxiongravitydilaton} in the spherically symmetric ansatz \eqref{eq:sphericalansatz}):
\begin{equation}
	\left\lbrace
	\begin{aligned}
		&\frac{2a\ddot{a}}{h}+\frac{\dot{a}^2}{h}-h-\frac{2a\dot{a}\dot{h}}{h^2}+\kappa a^2\left(\frac{\dot{\phi}^2}{2 h}+h V(\phi)\right)
-\frac{\kappa N^2 h}{ a^4}\,\e^{-\beta \phi\sqrt{\kappa}} =0\,,\\
		&\frac{\dot{a}^2}{h^2}-1=\frac{\kappa a^2}{3}\left(\frac{\dot{\phi}^2}{2 h^2}-V(\phi)\right)-
\frac{\kappa N^2}{3 a^4} \e^{-\beta \phi\sqrt{\kappa}}\,, \\
		&\ddot{\phi}+\left(\frac{3\dot{a}}{a}-\frac{\dot{h}}{h}\right)\dot{\phi}=h^2 \frac{\dd V}{\dd\phi}
-\frac{ \beta N^2 h^2\sqrt{\kappa}}{ a^6} \e^{-\beta \phi\sqrt{\kappa}}\, .
	\end{aligned}
	\right.\label{eq:eomAxionGravityDilaton}
\end{equation}
In the gauge $h\equiv 1$,  these equations simplify to
\begin{equation}
	\left\lbrace
	\begin{aligned}
		&2a\ddot{a}+\dot{a}^2-1+\kappa a^2\left(\frac{\dot{\phi}^2}{2}+V(\phi)\right)
-\frac{\kappa N^2}{ a^4} \e^{-\beta \phi\sqrt{\kappa}} =0\,,\\
		&\dot{a}^2-1=\frac{\kappa a^2}{3}\left(\frac{\dot{\phi}^2}{2}-V(\phi)\right)-
\frac{\kappa N^2}{3 a^4} \e^{-\beta \phi\sqrt{\kappa}}\,,\\
		&\ddot{\phi}+\frac{3\dot{a}}{a}\dot{\phi}=\frac{\dd V}{\dd\phi}
-\frac{ \beta N^2\sqrt{\kappa}}{ a^6} \e^{-\beta \phi\sqrt{\kappa}} \pkt
	\end{aligned}
	\right.\label{eq:fulleomh1}
\end{equation}

Using the trace of the first equation in \eqref{eq:eomfullaxiongravitydilaton}, the on-shell action of the wormhole solution can be easily calculated
\beq
S_\text{E}&=&\int\dd^4x\sqrt{g}\left(\frac{\e^{-\beta \phi\sqrt{\kappa}}}{6 f^2} H^2 - V(\phi)\right) \\
&=& 2 \pi^2 \int \dd\tau h a^3 \left(\frac{2 N^2 \e^{-\beta \phi\sqrt{\kappa}}}{ a^6}  - V(\phi) \right) \label{eq:action1}\kma
\eeq
where in the second line we used the spherically symmetric ansatz. Note that this expression for the on-shell action is equivalent to the action \eqref{eq:redact} upon using the constraint \eqref{eq:eomAxionGravityDilaton} and keeping the surface term.
From this expression of the action \eqref{eq:action1} we can conclude that for some potentials the wormhole action can become negative\footnote{This fact that the wormhole action could be negative for some potentials was already noticed in \cite{Shvedov:1996hb}.}. In section \ref{sec:axionscalar} wee will encounter numerical examples confirming this expectation.

The GS wormhole solution \cite{Giddings:1987cg} has $V=0$
and in the gauge $a=\tau$ can be written as \cite{Hebecker:2016dsw}
\be
a(\tau)=\tau\,,\quad h(\tau)= \left(1-\frac{a_0^4}{\tau^4}\right)^{-1/2},\quad \e^{\beta \phi(\tau)\sqrt{\kappa}}=
\frac{\kappa N^2}{3a_0^4} \cos^2\left[\frac{\beta}{\beta_c} \arccos (\frac{a_0^2}{\tau^2})\right],
\ee
where $\beta_c$ denotes the critical value of the dilaton coupling above which no solution exists, with
\be
a_0^4 = \frac{\kappa N^2}{3} \cos^2\left(\frac{\pi}{2}\frac{\beta}{\beta_c}\right),\qquad 0 \leq \beta < \beta_c = \frac{2\sqrt{2}}{\sqrt{3}} \pkt
\ee
The action of this wormhole can be easily calculated and reads
\be
 S^{(\beta)}_{GS}=2 \pi^2 \int_{a_0}^{\infty} \dd\tau\,\frac{2 N^2 \e^{-\beta \phi\sqrt{\kappa}}}{\tau^3 \sqrt{1-\frac{a_0^4}{\tau^4}}}
 = \frac{4 \sqrt{2} \pi^2 N}{\beta \sqrt{\kappa}} \sin(\frac{\pi}{2}\frac{\beta}{\beta_c})\,.
\ee
In the limit $\beta \to 0$, one gets a version of the GS solution without dilaton, whose action is
\be
S^{(0)}_{GS} = \frac{\sqrt{3} \pi^3 N}{ \sqrt{\kappa}} \pkt
\ee
More precisely what we are discussing here is the action of a semi-wormhole; to get an action for the whole wormhole with two asymptotic regions we should just multiply this result by a factor of two.


\subsection{Baby universe interpretation} \label{sec:bu}

Euclidean wormholes can be interpreted as tunnelling events leading to the creation of baby universes \cite{Giddings:1987cg,Coleman:1988cy}.
It was noted in \cite{Lavrelashvili:1988un} that GS wormholes are leading to the materialisation of baby universes
which are {\it contracting} after analytic continuation to Minkowski time.
Indeed, a regular wormhole at $\tau =0$ has finite size $a(0)=a_0 \neq 0 \kma$
and zero derivative $\dot{a}(0)= 0$ such that for small $\tau$ we can expand
\be
a(\tau)=a_0 + \frac{1}{2} a_2 \tau^2 + {\cal{O}}(\tau^4) \kma
\ee
where the coefficient $a_2 = \ddot{a}(0)$. After analytic continuation to Minkowski time $t=-i \tau$ we get
\be
a(t)=a_0 - \frac{1}{2} a_2 t^2 + {\cal{O}}(t^4) \pkt
\ee
Now it is clear that $a_2>0$ and $a_2 <0$ correspond respectively to contracting and expanding small universes.
The GS wormhole obviously has $a_2=\ddot{a}(0)>0$, since the neck of the wormhole is a minimum of $a(\tau)$.
Instead, a wormhole leading to an {\it expanding} baby universe should have  $a_2=\ddot{a}(0)<0$, {\it i.e.} the ``neck'' of such a wormhole should be a local maximum. By combining the first two equations of \eqref{eq:fulleomh1}, one can see that the axion charge is required so that there can be a local minimum of the scale factor somewhere in the geometry.


\subsection{Initial conditions} \label{sec:ic}

The variational problem following from the reduced action \eqref{eq:redact} is well-defined only if the following boundary terms vanish:
\begin{equation} \left[\frac{3a^2}{h\kappa}\delta\dot{a}\right]_0^{\tau_f}=0\quad\text{and}\quad\left[\frac{a^3\dot{\phi}}{h}\delta\phi\right]_0^{\tau_f}=0\,.
\end{equation}
This is realised at the wormhole neck $\tau=0$ for the initial conditions $\dot{a}(0)=0$ and $\dot{\phi}(0)=0$, and in the asymptotic future for the conditions $\dot{a}(\tau_f)=1$ and $\phi(\tau_f)=0$, which imply that the asymptotic future is the flat Euclidean spacetime. This explains why the Gibbons-Hawking-York term must not be added to the action in this case, as it would suppress the boundary term in \eqref{eq:redact} and consequently would imply that the value of the scale factor $a$ must be fixed at the boundary instead of $\dot{a}$; yet fixing $\dot{a}$ is precisely what allows us to reach flat Euclidean spacetime in the asymptotic future as well as impose the wormhole constraint $\dot{a}(0)=0$ at the origin.

On the classical solution, we must also specify the initial values of the scale factor and scalar field. The value of the scalar field $\phi(0)=\phi_0$ is a free parameter, while the throat size, $a(0)=a_0$, is determined by the Friedmann constraint at $\tau=0$:
\begin{align}
	&1=\frac{\kappa}{3}\left(a_0^2V(\phi_0)+\frac{Q^2}{a_0^4}\right)\,,\label{eq:friedmann}\\
	\Leftrightarrow\ &\frac{\kappa}{3}V(\phi_0)x^3-x^2+\frac{\kappa Q^2}{3}=0\,,\quad x=a_0^2 \kma \label{eq:a0_constraint}
\end{align}
where we defined
\be
Q^2 = N^2  \e^{-\beta \phi_0 \sqrt{\kappa}}\,.
\ee
The discriminant of the cubic equation \eqref{eq:a0_constraint} is $\displaystyle\Delta=\frac{\kappa Q^2}{3}\left(4-\kappa^3Q^2V(\phi_0)^2\right)$. When $\Delta>0$, there are three real solutions for $x$, while when $\Delta<0$, there are one real and two complex solutions for $x$. For $\Delta>0$ we may equivalently require
\begin{equation}
	\frac{2}{Q}>\kappa^{3/2}V(\phi_0)\,.\label{eq:1stineq}
\end{equation}
The three solutions can be constructed as follows. Let us define an angle $\theta\in [0,\pi]$ such that
\begin{equation}
 \cos {\theta}= 1-\frac{1}{2}\kappa^3 Q^2 V^2(\phi_0) \,,\quad\theta= \arccos (1-\frac{1}{2}\kappa^3 Q^2 V^2(\phi_0))\,. \label{eq:thetadef}
\end{equation}
Then the solutions to Eq. \eqref{eq:a0_constraint} are given by
\begin{equation}
	x_j=\frac{1}{\kappa V(\phi_0)}\left(1+2\cos(\frac{\theta-2\pi\cdot j}{3}
	)\right)\quad\text{for}\ j=0,1,2\,. \label{eq:roots}
\end{equation}
Whatever the value of $\theta\in(0,\pi)$, if $V(\phi_0)>0$ there are always two of these solutions which are positive $(j=0,1)$ and one which is negative $(j=2)$. From the two positive solutions for $x$, we get four real solutions for $a_0$, two positive and two negative.
The largest positive solution for $a_0$ satisfies
\begin{equation}
	\sqrt{\frac{2}{\kappa V(\phi_0)}}\,<a_0^{(1)}<\,\sqrt{\frac{3}{\kappa V(\phi_0)}}\,,
\end{equation}
so combining the left inequality with \eqref{eq:1stineq}, we find $\left(a_0^{(1)}\right)^2>\kappa^{1/2} Q\,.$
Therefore
\begin{equation}
	\ddot{a}(0)=-\frac{1}{a_0^{(1)}}\left(1-\frac{\kappa Q^2}{\left(a_0^{(1)}\right)^4}\right)<0\,,
\end{equation}
{\it i.e.} $\displaystyle\frac{\dd^2 a}{\dd t^2}(0)=-\ddot{a}(0)>0$, in other words the solution with largest $a_0$ corresponds to a ``wineglass'' wormhole leading to an expanding baby universe. This is the $j=0$ solution. Assuming $y=\kappa^3Q^2V(\phi_0)^2 \ll 1$, we find that $\theta\simeq\pi$ so that
$\displaystyle a_0^{(1)}\simeq\sqrt{\frac{3}{\kappa V(\phi_0)}}$, which implies that the throat size is given by the Hubble scale. Meanwhile, the smaller ($j=1$) root in \eqref{eq:roots} leads to wormholes of GS type, with the scale factor being a local minimum at the origin.

For $\Delta<0$, the equation \eqref{eq:a0_constraint} only possesses one real root, which reads:
\begin{equation}
	x_0=\frac{1}{\kappa V(\phi_0)}\left[1-2\cosh(\frac{1}{3}\cosh[-1](\frac{\kappa^3Q^2V(\phi_0)^2}{2}-1))\right].
\end{equation}
Because $\Delta<0$, the arc-hyperbolic cosine in this expression is real, so $x_0$ is real and negative. The corresponding solution for $a_0$ is thus imaginary and can be discarded.

Therefore, $\Delta>0$ is a necessary (but not sufficient) condition on the parameter space for obtaining a wormhole solution.


\section{Axion-dilaton wormholes}

\begin{figure}[h]
	\centering
	\begin{subfigure}{0.49\linewidth}
		\includegraphics[width=6cm]{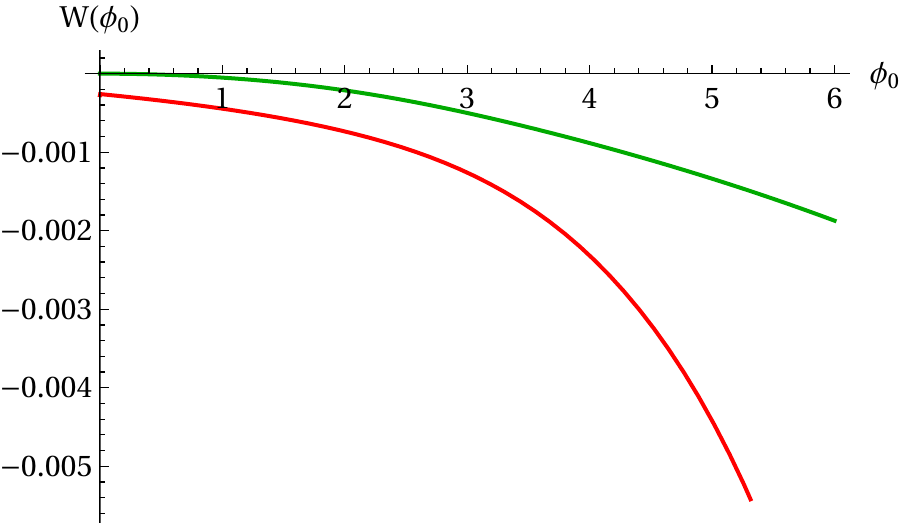}
        \caption{Monotonic effective potential.}\label{fig:W_a}
	\end{subfigure}
	\begin{subfigure}{0.49\linewidth}
		\includegraphics[width=6cm]{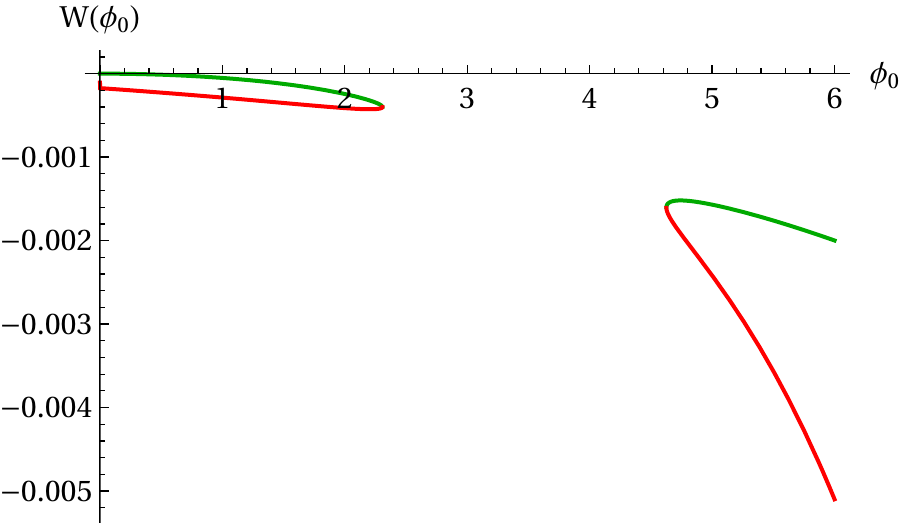}
        \caption{Non-monotonic effective potential.}\label{fig:W_b}
	\end{subfigure}
	\caption{\small Effective potential $W(\phi)$ as a function of $\phi_0$ for axion charge $N=20000$ (left) and $N=30000$ (right).
The other parameters are the same for both plots: $m=0.01$ and $\beta = 1.2$. The green line corresponds to the large root in \eqref{eq:roots}, while the red line corresponds to the smaller positive root. The plot on the right contains a gap, which is  caused by there existing no real solutions to the cubic equation \eqref{eq:a0_constraint} in that range.}\label{fig:W}
\end{figure}

After these preliminaries, we are ready to search for wormhole solutions. We will start with the case of the axion-dilaton-gravity system, where we assume the dilaton to be massive, that is to say we choose the potential
\be
V(\phi) = \frac{1}{2} m^2 \phi^2 \kma
\ee
where  $m$ is the dilaton mass. The equations of motion \eqref{eq:fulleomh1} in the gauge $h=1$ now read
\begin{equation}
	\left\lbrace
	\begin{aligned}
		&2a\ddot{a}+\dot{a}^2-1+\kappa a^2\left(\frac{\dot{\phi}^2}{2}+\frac{1}{2} m^2 \phi^2\right)
		-\frac{\kappa N^2}{ a^4} \e^{-\beta \phi\sqrt{\kappa}} =0\quad\text{(acceleration equation)}\,,\\
		&\dot{a}^2-1=\frac{\kappa a^2}{3}\left(\frac{\dot{\phi}^2}{2}-\frac{1}{2} m^2 \phi^2\right)-
		\frac{\kappa N^2}{3 a^4} \e^{-\beta \phi\sqrt{\kappa}}
		\quad\text{(Friedmann constraint)}\,,\\
		&\ddot{\phi}+\frac{3\dot{a}}{a}\dot{\phi}= m^2 \phi
		-\frac{ \beta N^2\sqrt{\kappa}}{ a^6} \e^{-\beta \phi\sqrt{\kappa}}\quad\text{(dilaton equation)}\pkt
	\end{aligned}
	\right.\label{eq:dilaton}
\end{equation}
The dilaton equation in \eqref{eq:dilaton} possesses a mechanical analogy as
the motion of a \textit{particle} $\phi(\tau)$ in an \textit{effective potential} $W(\phi)$:
\beq
W(\phi)&=&- V(\phi) - \frac{N^2}{a^6} \e^{-\beta \phi\sqrt{\kappa}} \kma \\
\frac{\dd W(\phi)}{\dd\phi}&=&- m^2 \phi + \frac{ \beta N^2\sqrt{\kappa}}{a^6} \e^{-\beta \phi\sqrt{\kappa}} \pkt
\eeq
We see that the fate of the particle released at some point $\phi(0)=\phi_0 >0$ with zero velocity $\dot{\phi}(0)=0$ depends on the sign of $W_{,\phi}$ at this point. Depending on which term in the potential $W(\phi)$ dominates at this point,
the particle either starts to move to the right (increasing $\phi$) or to the left (decreasing $\phi$).
Since we want to obtain an asymptotically flat geometry, for $\tau \to \infty$ the dilaton field should eventually settle at its
vacuum value $\phi = 0$. The shape of the effective potential crucially depends on the axion charge $N$, and also on which root is chosen in \eqref{eq:roots} -- see Fig.~\ref{fig:W} for an illustration. One feature that becomes immediately clear upon inspection of the effective potential is that solutions in which the dilaton monotonically rolls down its potential can only exist in very limited parameter ranges, precisely in the regions near the gap seen in the right hand plot. This property is useful in the numerical search for solutions, to which we now turn.

\subsection{Generalisations of Giddings-Strominger wormholes}

We will first concentrate on generalisations of the GS solution (by which we mean wormholes leading to the nucleation of contracting baby universes), in the presence of a massive dilaton. This type of solution was already discussed in the recent work \cite{Andriolo:2022rxc}, though our search is more comprehensive and we find numerous additional solutions here. In \cite{Andriolo:2022rxc}, it was shown that solutions only depend on the combination of parameters $m^2q/f$ and $\beta$, a result which follows from the possible field rescalings which we review in appendix \ref{sec:scaling}. The authors of \cite{Andriolo:2022rxc} fixed $q/f=\sqrt{2}\Leftrightarrow N=1$ and varied $m$. Instead, in our numerical analysis we have fixed $m=10^{-2}$ and varied $N$. The two procedures are exactly equivalent, note however that in \cite{Andriolo:2022rxc} they use a different, non-canonical definition of the dilaton potential so that their mass is rescaled, $m_\text{there}=m_\text{here}/\sqrt{2},$ compared to ours.

\begin{figure}
	\includegraphics[width=0.45\textwidth]{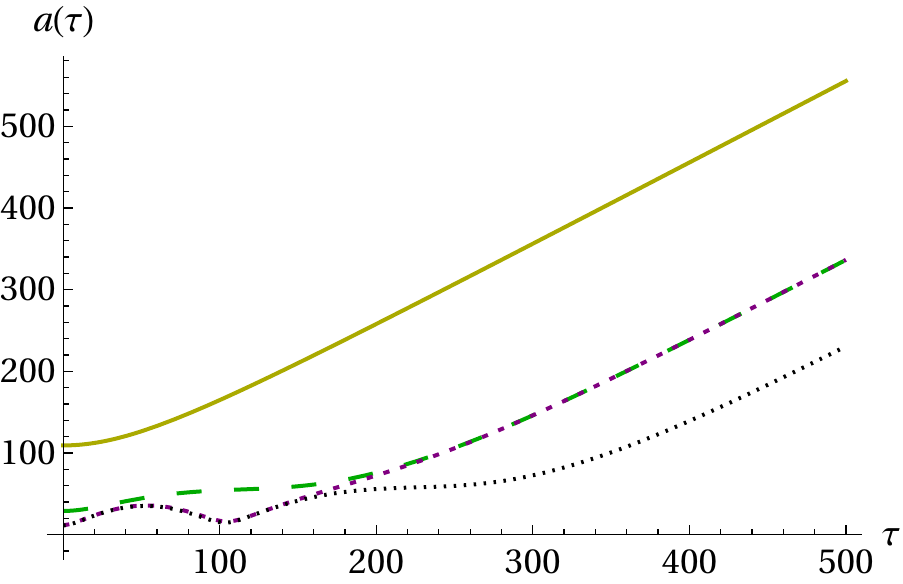}
	\includegraphics[width=0.45\textwidth]{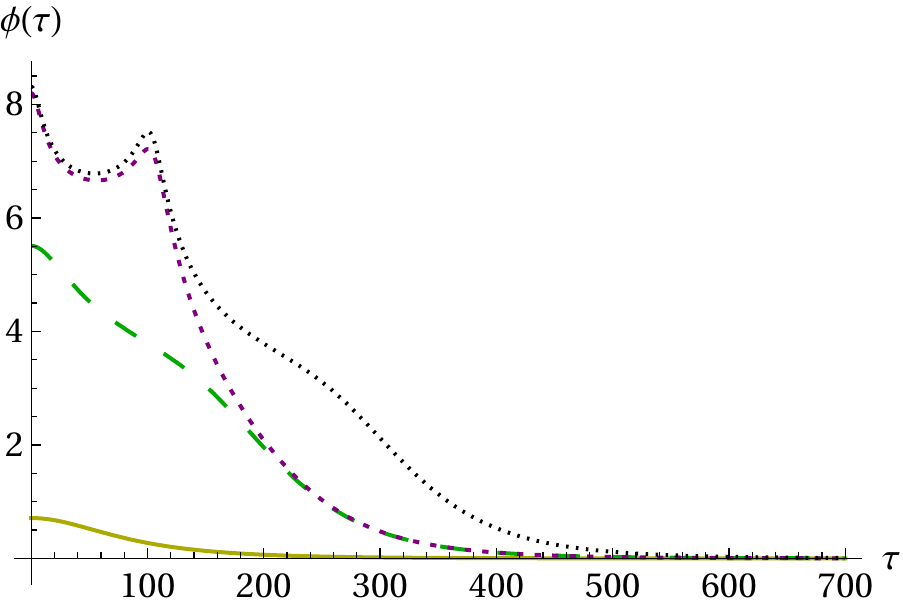}
	\caption{Wormhole solutions with a massive dilaton, with the scale factor shown on the left and the dilaton evolution on the right. All solutions have $\kappa=1,$ $\beta=1.2,$ $N=30000,$ $m=0.01.$ The individual solutions are characterised by the initial value of the dilaton, given respectively by the values $\phi_0=0.7118165858, 5.5075291704, 8.1964321797, 8.3116654157$ (we indicate a number of significant digits such that the action can be determined to better than percent level accuracy). Solutions with larger $\phi_0$ display a more intricate field evolution, containing oscillations of the fields.}\label{fig:GS12}
\end{figure}

In looking for solutions, we search for domains of the initial dilaton value $\phi_0$ in which we can identify an over-/undershooting behaviour of the solution. By this we mean that we look for values of $\phi_0$ such that small changes cause the asymptotic value of the dilaton to switch between running off to plus or minus infinity. Then we can infer, by continuity, that somewhere in between a solution must exist in which the dilaton approaches $\phi=0$ asymptotically. We may hone in on the actual solution to the desired level of accuracy (which is typically at the level of about $20$ significant digits, sometimes more) with an optimisation algorithm. Here a Newtonian algorithm suffices. In finding appropriate initial ranges of $\phi_0,$ we take guidance from the effective potential, as discussed above.

\begin{figure}
	\includegraphics[width=0.45\textwidth]{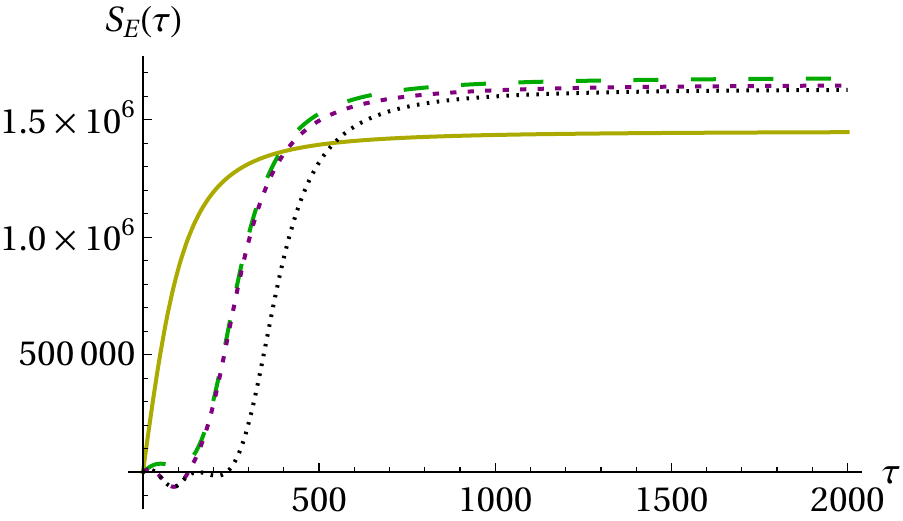}
	\includegraphics[width=0.45\textwidth]{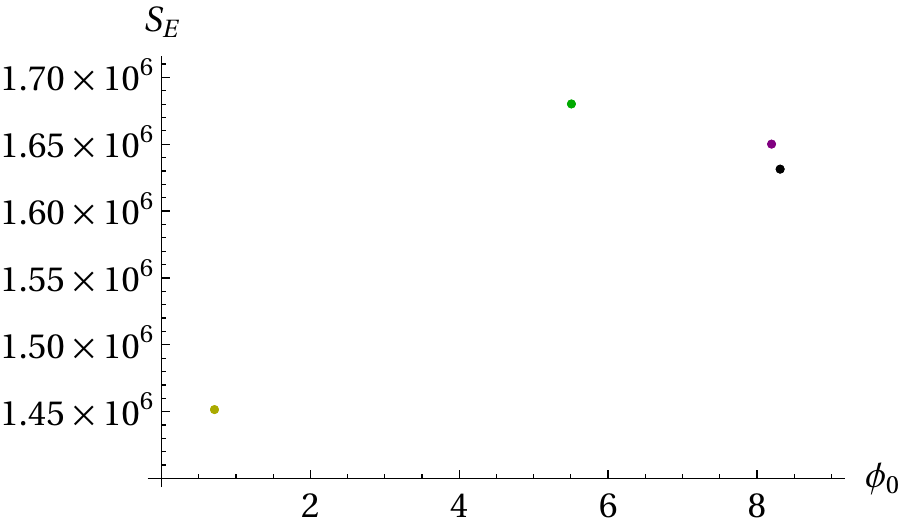}
	\caption{The Euclidean action, as a function of $\tau$ (left plot) and a graph with the asymptotic values (right plot), for the solutions shown in Fig.~\ref{fig:GS12}. Intriguingly, the action is not monotonic in $\phi_0,$ but starts decreasing as more oscillations are added.}\label{fig:GS12S}
\end{figure}

The first examples of solutions are shown in Fig.~\ref{fig:GS12}. These solutions all have the same dilaton coupling $\beta=1.2,$ and the same axion charge $N=30000,$ so that $m^2N=3.$ For all solutions, the origin $a(0)$ represents a local minimum of the scale factor, which is why these wormholes would nucleate contracting universes. In \cite{Andriolo:2022rxc} only solutions analogous to the first, monotonic solution with $\phi_0 \approx 0.71$ were found. Remarkably, with increasing $\phi_0,$ we find additional solutions in which the field evolutions become more intricate, and oscillations develop. It is noteworthy that both the dilaton and the scale factor develop these oscillations (in contrast to the oscillating bounces of \cite{Hackworth:2004xb}, in which only the scalar field oscillates). In the figure we show four examples, although we did not encounter any obstruction when searching for solutions with ever larger $\phi_0$ and more numerous oscillations -- we suspect that there exists no upper limit to the number of oscillations.

The Euclidean action of these solutions is shown in Fig.~\ref{fig:GS12S}, both as a function of the radial coordinate $\tau$ and in terms of the final asymptotic values. One can see that the action integral only receives significant contributions as long as the dilaton is evolving noticeably. Once the dilaton settles in its potential minimum and the geometry approaches that of flat space, the action stabilises. To obtain the asymptotic value, we have added an analytically calculable remainder (as explained in appendix \ref{appendix:action_correction}), which ensures that the end result is trustworthy up to a known level of precision. The most obvious feature is that the action is positive, in line with the interpretation of these solutions as mediating the nucleation of baby universe in a tunnelling-type event. The second solution, containing an inflection point in the scale factor, has a higher action. However, what is truly surprising, is that for the solutions with additional features, in particular the presence of one or two additional minima in $a(\tau),$ the action decreases again. Assuming the probability of nucleation per unit four-volume to be given approximately by $e^{-2S_E/\hbar},$ one would conclude that solutions with additional oscillations are more likely to occur than those with fewer such features. This surprising property requires further explanation. Notice however, for now, that the simplest, monotonic solution appears to be the most likely overall, though in the absence of analytic expressions for these solutions we cannot be sure what happens to the action in the limit of infinite numbers of field oscillations. In particular, we cannot assess whether in this limit the action would remain above that of the lowest $\phi_0$ solution.

\begin{figure}
	\includegraphics[width=0.45\textwidth]{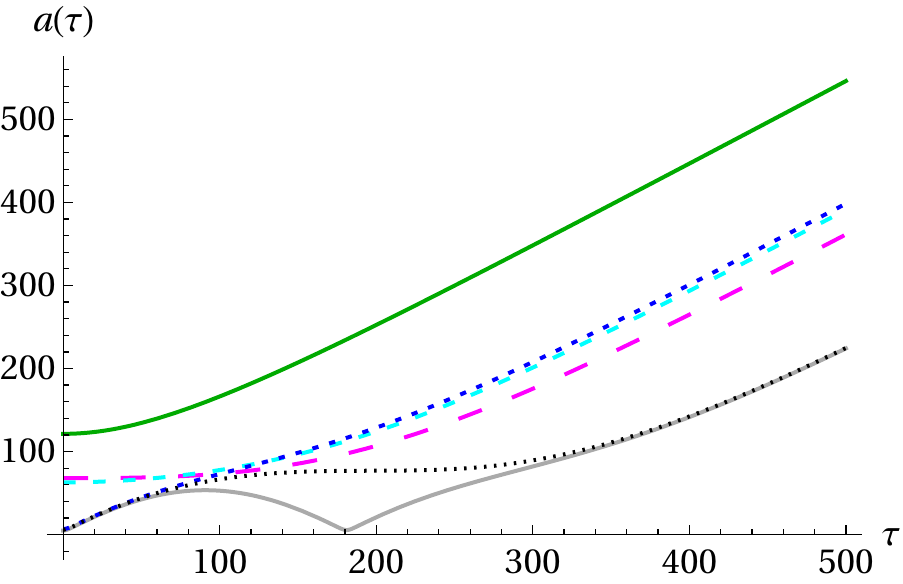}
	\includegraphics[width=0.45\textwidth]{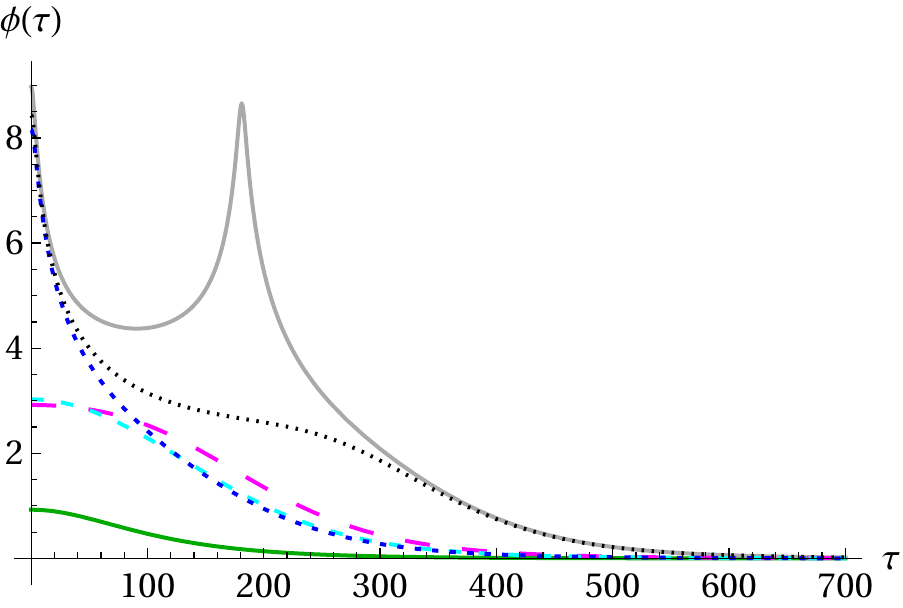}
	\caption{Wormhole solutions with a massive dilaton, with the scale factor shown on the left and the dilaton evolution on the right. All solutions have $\kappa=1,$ $\beta=1.58,$ $N=47089,$ $m=0.01.$ The initial dilaton values are $\phi_0=0.9267658893,$ $2.9202136114,$ $3.0261054894,$ $8.1578681214,$ $8.4314038628,$ $8.9744628254.$ Solutions with larger $\phi_0$ again display oscillations of the fields.}\label{fig:GS158}
\end{figure}

\begin{figure}
	\includegraphics[width=0.45\textwidth]{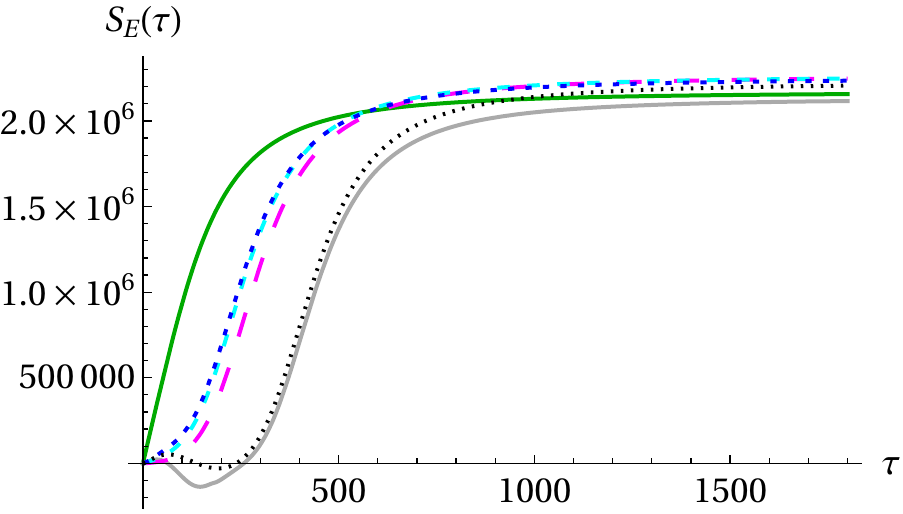}
	\includegraphics[width=0.45\textwidth]{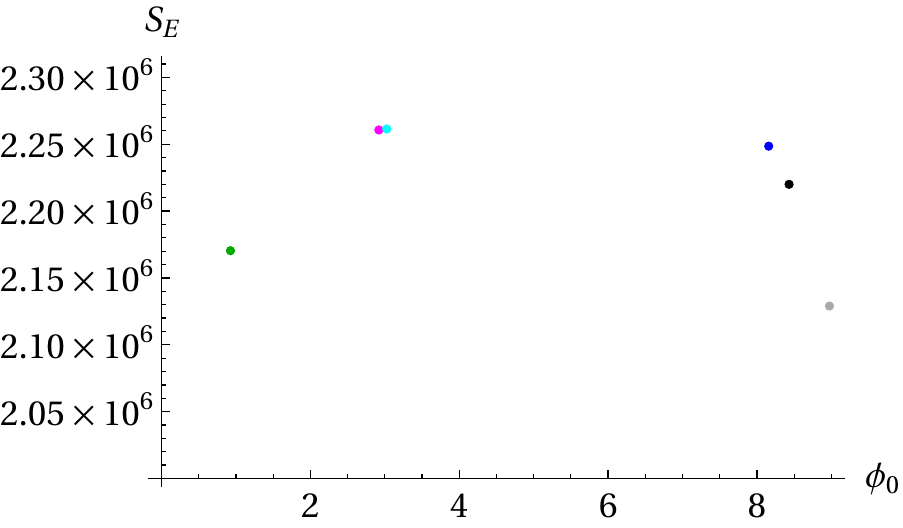}
	\caption{The Euclidean action, as a function of $\tau$ (left plot) and a graph with the asymptotic values (right plot), for the solutions shown in Fig.~\ref{fig:GS158}. Again, the action is not monotonic in $\phi_0,$ but starts decreasing as more oscillations are added.}\label{fig:GS158S}
\end{figure}

We can also look at solutions at a larger value of the dilaton coupling -- Fig.~\ref{fig:GS158} shows an example with $\beta=1.58$ and with charge $N=47089$ (the reason for choosing this peculiar looking number will become clear below). At these parameter values we find a total of five GS-type solutions, containing a single minimum in the scale factor, that is to say only the minimum at the origin. The figure then also contains a solution with two minima, at an even larger dilaton value $\phi_0.$ The action of these solutions is shown in Fig.~\ref{fig:GS158S}. We notice the same feature as before, namely that the action first increases with the complexity of the solutions (and with increasing $\phi_0$) but then starts to decrease again. This time the effect is even more pronounced, and the solution with two minima of the scale factor already has a Euclidean action that lies below that of the lowest $\phi_0$ solution. We will discuss this puzzling feature further in later parts of the paper.

\begin{figure}[h!]
	\includegraphics{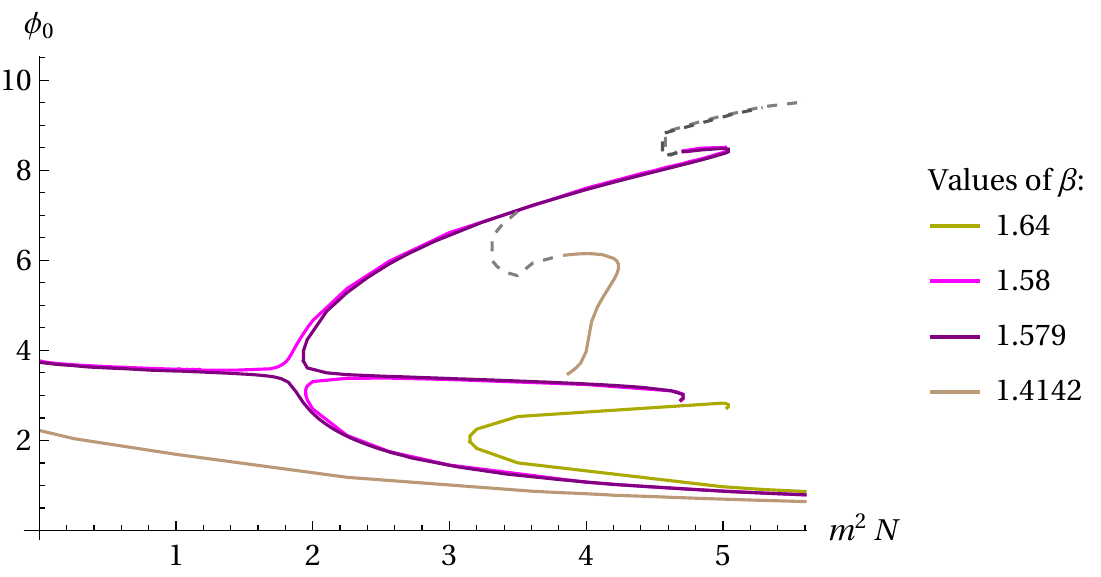}\caption{Branch structure of the generalised GS-type solutions with at most one extra minimum of the scale factor, for four representative values of the dilaton coupling. A full description is provided in the main text.}\label{fig:branchstructureGSred}
\end{figure}

\begin{figure}[h!]
	\includegraphics{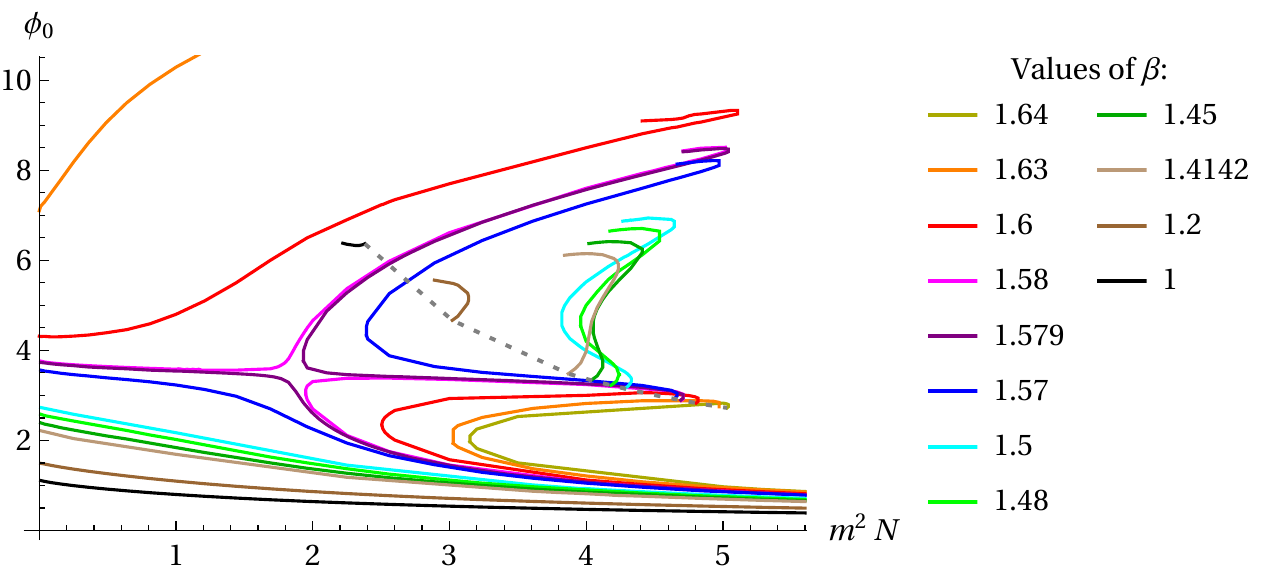}\caption{A more complete version of the plot shown in Fig.~\ref{fig:branchstructureGSred}, containing curves at additional values of the dilaton coupling $\beta.$ The inversion of the branch structure above $\beta_i$ ({\it i.e.} for the curves with $\beta \geq 1.58$) is clearly visible. See the main text for a detailed analysis.}\label{fig:branchstructureGS}
\end{figure}

The existence of these generalised GS-solutions depends on the parameters $m^2 N$ and $\beta,$ as well as on the initial dilaton values $\phi_0,$ in a remarkably intricate way. We illustrate this first with the simplified plot in Fig.~\ref{fig:branchstructureGSred}, and then with the more complete (but at first sight somewhat bewildering) plot in Fig.~\ref{fig:branchstructureGS}. Let us describe the simplified plot first, starting with the light brown curve corresponding to $\beta = \sqrt{2} \approx 1.4142.$ The lowest curve on the plot shows that wormhole solutions exist for all values of $m^2N,$ {\it i.e.} also in the limit of vanishing mass. This is expected, as this value of $\beta$ lies below the critical GS value $\beta_c =\sqrt{8/3} \approx 1.633.$ However, an additional branch of solutions exists, starting around $m^2 N \approx 4$ and $\phi_0 \approx 4.$ The starting value of this new branch is not arbitrary: it corresponds to the onset of the ``gap'' in the effective potential $W(\phi),$ as explained in and above Fig.~\ref{fig:W}. Along this new branch lie the solutions with inflection points and oscillations in the fields. In fact, as one moves along the branch, the solutions become progressively more complicated. The solid light brown curve includes all solutions (with $\beta=\sqrt{2}$) that have a single minimum in the scale factor. The dashed extension of the curve indicates that further solutions, with two and more minima of the scale factor, exist as one moves in the broad direction of larger $\phi_0$ values. In fact, as far as we can tell by our numerical investigations, the curve keeps moving up in a zig-zag manner, without apparent limit. 

This behaviour is analogous for all values of $\beta$ below a certain $\beta_i,$ with $1.579 < \beta_i < 1.58,$ at which value an ``inversion'' of the branch structure takes place. Indeed, the pink curve ($\beta=1.58$) that links to the vertical axis at $m=0$ now moves up to larger $\phi_0$ values around $m^2 N \approx 2,$ and the new branch of solutions appears below, at smaller $\phi_0.$ Now it is the original branch that continues upwards in a zig-zag manner, indicated by the dashed extension of the solid line, leading to ever more involved field evolutions. By contrast, the new branch that appears around $m^2 N\approx 2$ has a bifurcating pattern. Following it along the larger $\phi_0$ values, it has an end point at $m^2 N \approx 4.8$ and $\phi_0 \approx 3,$ corresponding to the (dis)appearance of the gap in the effective potential $W(\phi).$ Following this curve in the other direction, one finds solutions with ever larger $m^2 N,$ again presumably without limiting $m^2 N$ value. The solutions shown in Fig.~\ref{fig:GS158} present a vertical slice through these curves, and from Fig.~\ref{fig:branchstructureGSred} one can see that choosing $m^2 N$ to lie a little below $4.8$ allows one to pick up the maximal variety of solutions.

If we increase the dilaton coupling $\beta$ further, then on the vertical axis, at zero dilaton mass, the required $\phi_0$ values keep increasing and tend to infinity as the critical value $\beta_c =\sqrt{8/3} \approx 1.633$ is reached. This is the limit found by Giddings and Strominger for the massless case. Beyond this value, the line that connects to the vertical axis thus disappears. Consider for instance the light green curve at $\beta=1.64$ in Fig.~\ref{fig:branchstructureGSred}. At this value, only the new bifurcating branch of the type described above exists. For this value of $\beta,$ it appears around $m^2 N \approx 3.2,$ and this implies that GS-type wormhole solutions only exist for masses/charges above this value.

We can now take a look at the more complete plot shown in Fig.~\ref{fig:branchstructureGS}. One can make out very well the similarities between all curves with $\beta < \beta_i,$ and the inverted structure at larger coupling. For completeness, one should picture the upper branches (for $\beta<\beta_i$) as continuing in the described zig-zag pattern to larger $\phi_0$ values. The dashed line in the figure corresponds to the combination of $m^2 N$ and $\phi_0$ values at which the gap in the effective potential appears, {\it i.e.} the dashed line is the locus of end points of the new branches. The fact that an inversion of the branch structure occurs as $\beta$ is increased is made plausible by the fact that the curve that connects to zero mass must disappear entirely once $\beta$ surpasses $\beta_c.$ However, this argument is not sufficient to determine the precise value $\beta_i$ of the dilaton coupling at which the inversion occurs. This is an open problem, the resolution of which will likely require at least a partially analytic understanding of these numerically found solutions.

\begin{figure}[h!]
	\includegraphics[width=0.4\textwidth]{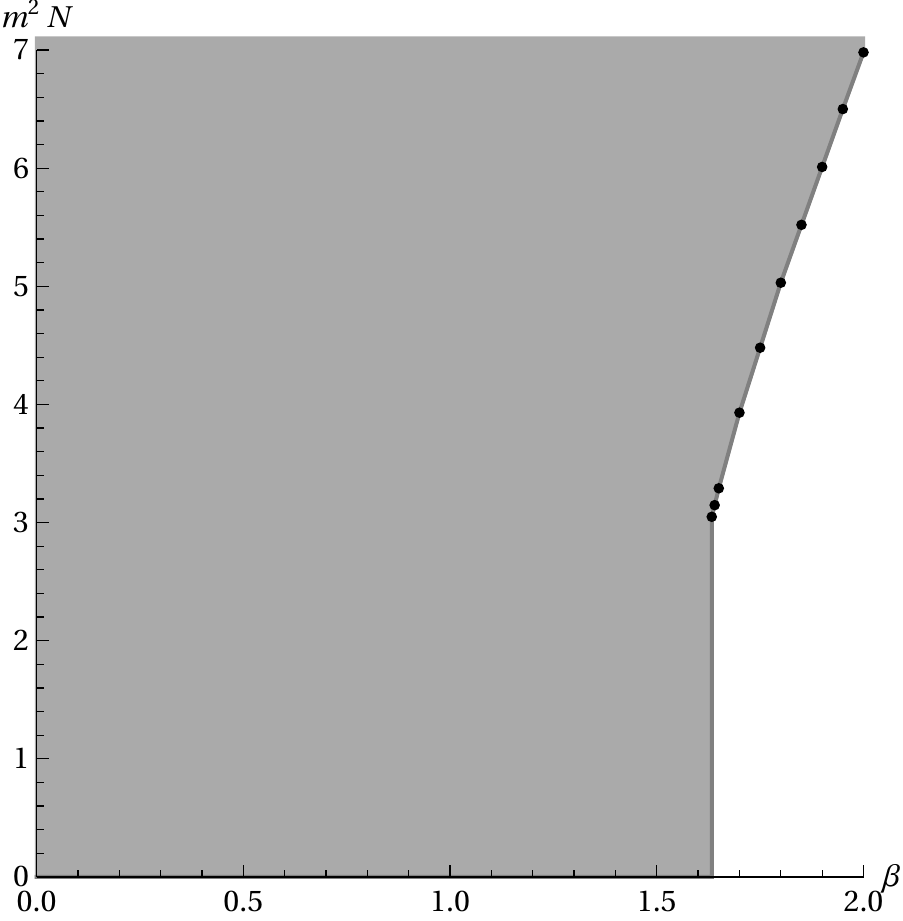}\caption{Minimal mass/charge $m^2 N$ as a function of $\beta$ for generalised GS-type solutions.}\label{fig:betam2N}
\end{figure}

Connected to the above point is the realisation that GS-type wormholes in fact exist at $\beta>\beta_c,$ but only with a sufficiently large mass and charge. This fact was already discovered in \cite{Andriolo:2022rxc}. Fig.~\ref{fig:betam2N} makes this notion more precise, by plotting the required minimum values of $m^2 N$ as a function of $\beta.$ The discontinuous jump at $\beta_c$ is precisely due to the appearance of new branches disconnected from the vertical $m^2 N=0$ axis. The obvious implication of this plot is that the required $m^2 N$ value becomes ever larger as the dilaton coupling is increased, with the approximate linear relation
\begin{align}
(m^2 N)_{min} \approx 10.8 \, \beta - 14.6\,,
\end{align}
valid for $\beta \geq \beta_c$.

\subsection{Wormholes leading to expanding baby universes}

The same axion-dilaton-gravity theory with massive dilaton admits another type of wormhole solution, in which the origin $a(0)$ corresponds to a local maximum of the scale factor. As described in section \ref{sec:bu}, if we interpret such solutions as mediating the nucleation of baby universes, then this feature leads to baby universes that are expanding (in Lorentzian time), a feature that may allow them to become large and be long lived. We will loosely refer to such solutions as {\it expanding wormholes}. The methods used to find such expanding wormholes are the same as those described in the previous section, except that in fixing initial conditions, we must always choose the largest root of the cubic equation \eqref{eq:a0_constraint}, as explained in section \ref{sec:ic}. 

\begin{figure}[h!]
	\includegraphics[width=0.31\textwidth]{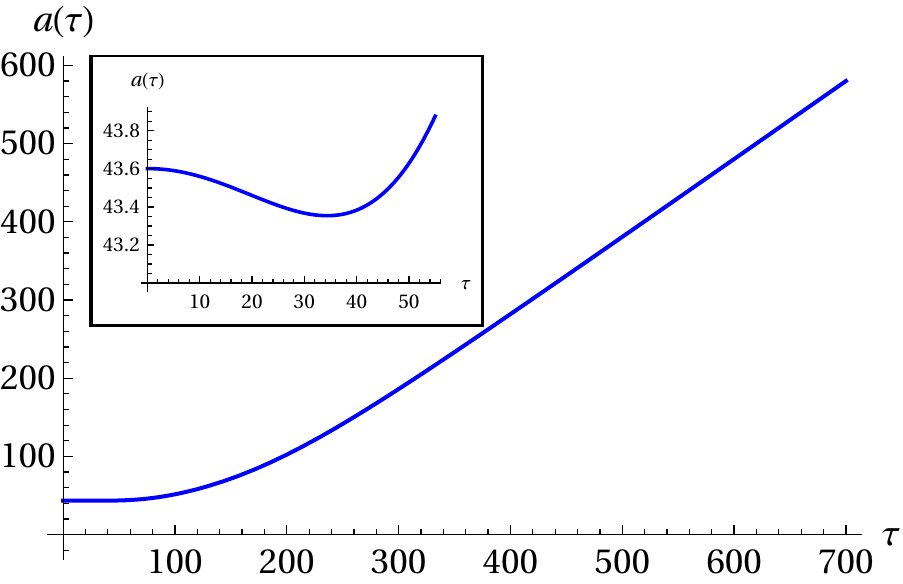}
	\includegraphics[width=0.31\textwidth]{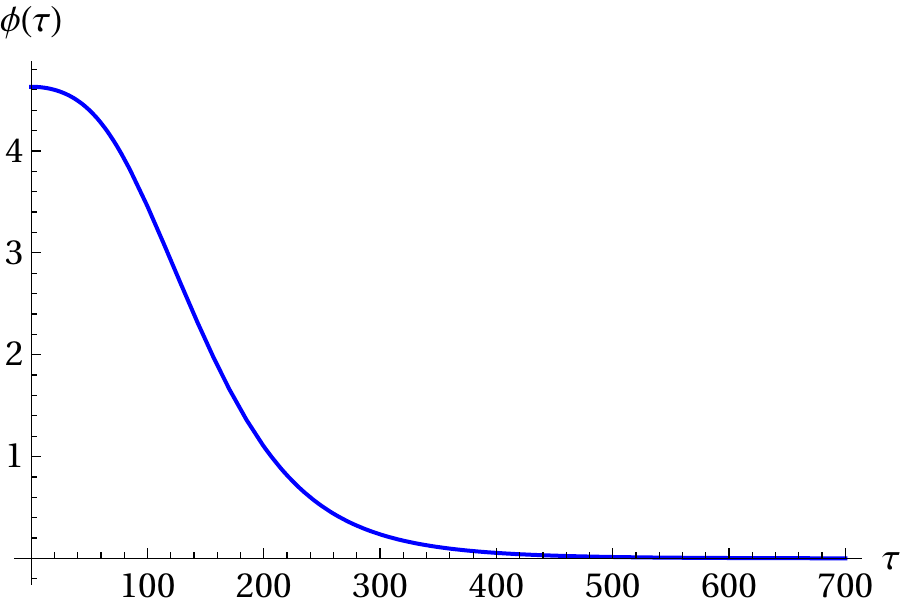}
	\includegraphics[width=0.31\textwidth]{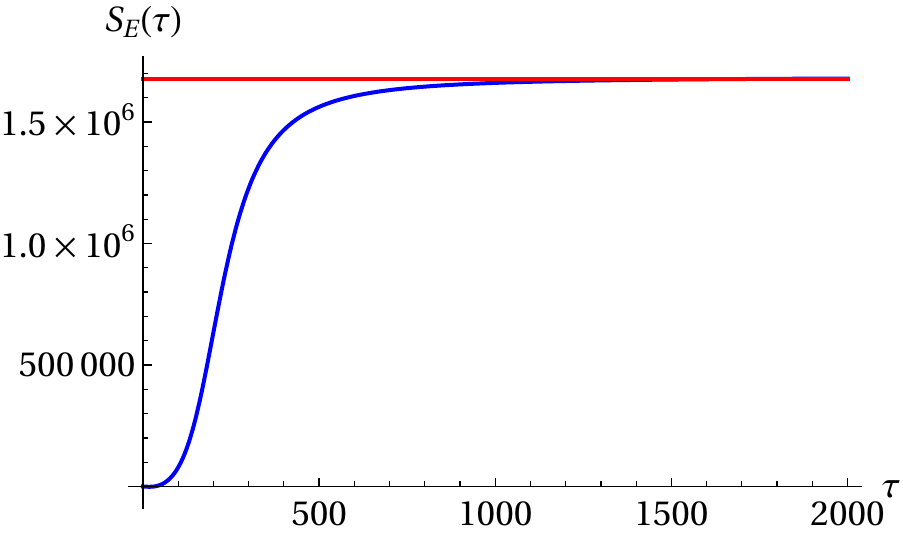}
	\caption{An example of a wormhole leading to an expanding universe upon analytic continuation. For this solution, the dilaton rolls down the potential monotonically. Shown are the scale factor (left), dilaton (middle) and Euclidean action (right). The red line represents the value of the final Euclidean action when taking the analytic remainder into account (see Appendix \ref{appendix:action_correction}). The parameter values are $m=0.01,$ $\beta = 1.2,$ $N=30000$ and the initial dilaton value is $\phi_0=4.6297956230.$} \label{fig:rolldown}
\end{figure}

A first example of such a wormhole is shown in Fig.~\ref{fig:rolldown}. One can only make out the local maximum of the scale factor at the origin by zooming in, see the inset in the left panel in the figure. For this solution, the dilaton rolls down monotonically in its potential, asymptotically settling at its minimum. The Euclidean action is positive, in line with the interpretation of this solution as mediating a tunnelling event.

\begin{figure}
	\includegraphics[width=0.45\textwidth]{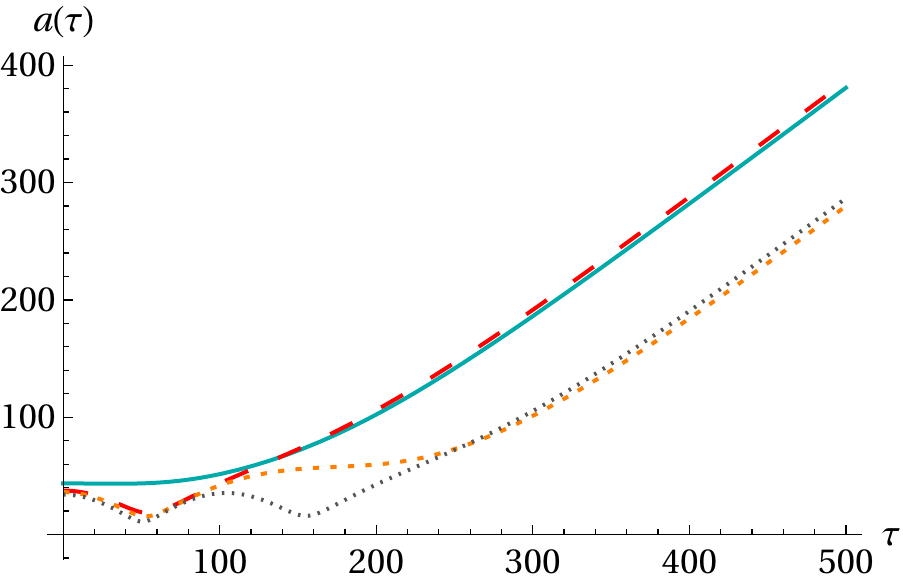}
	\includegraphics[width=0.45\textwidth]{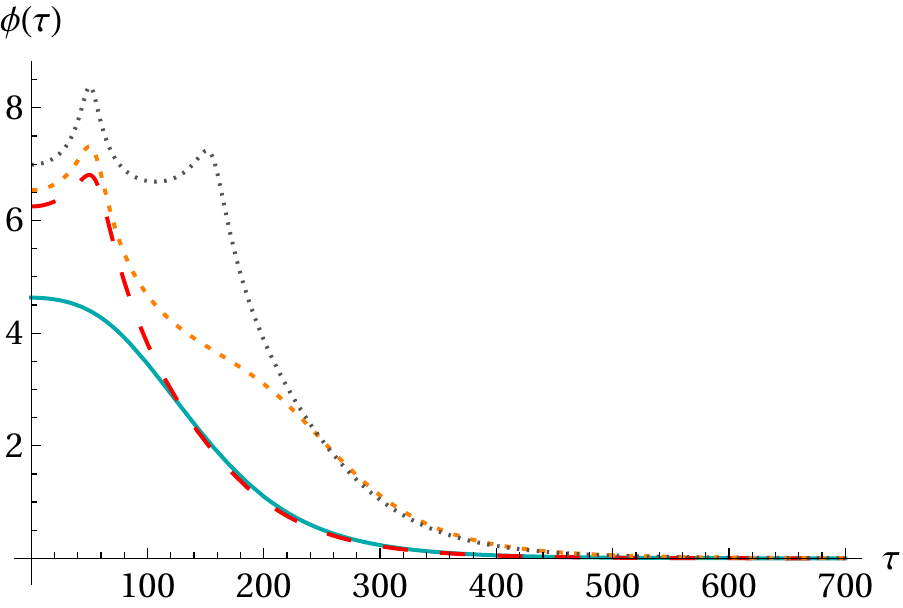}
	\caption{Expanding wormhole solutions with a massive dilaton, with the scale factor shown on the left and the dilaton evolution on the right. All solutions have $\kappa=1,$ $\beta=1.2,$ $N=30000,$ $m=0.01.$ The individual solutions are characterised by the initial value of the dilaton, given respectively by the values $\phi_0=4.6297956230, 6.2498081147, 6.5411315634, 6.9914512133.$ Solutions with larger $\phi_0$ display a more intricate field evolution, containing oscillations of the fields.}\label{fig:GS12exp}
\end{figure}

With the same parameters, there exist additional solutions at larger initial dilaton values, see Fig.~\ref{fig:GS12exp}. These solutions are reminiscent of the solutions with oscillations in the fields, discovered also for GS-type wormholes in the previous section. A distinction here is that in these more involved solutions the dilaton first runs up its potential, before turning around and (perhaps after several additional oscillations) eventually settling in its potential minimum.  Associated with these oscillations of the dilaton is an additional ``breathing'' behaviour of the scale factor, which may expand and shrink alternately before eventually tending to flat space at large radii. 

\begin{figure}
	\includegraphics[width=0.45\textwidth]{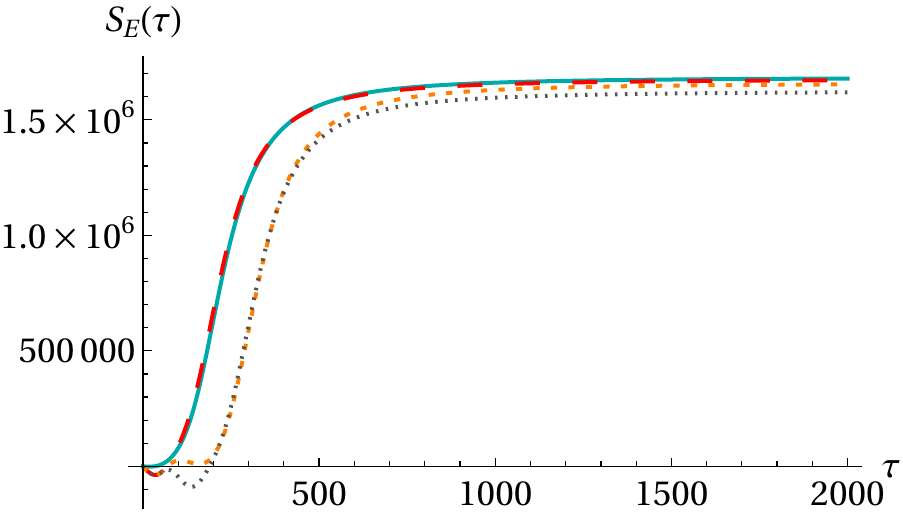}
	\includegraphics[width=0.45\textwidth]{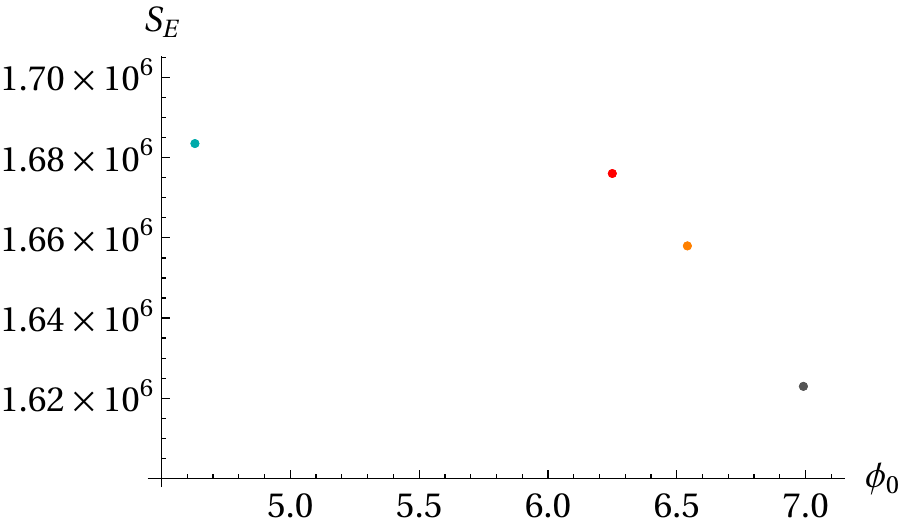}
	\caption{The Euclidean action, as a function of $\tau$ (left plot) and a graph with the asymptotic values (right plot), for the solutions shown in Fig.~\ref{fig:GS12exp}. Surprisingly, the action decreases as the field evolutions become more involved.}\label{fig:GS12expS}
\end{figure}

\begin{figure}
	\includegraphics[width=0.45\textwidth]{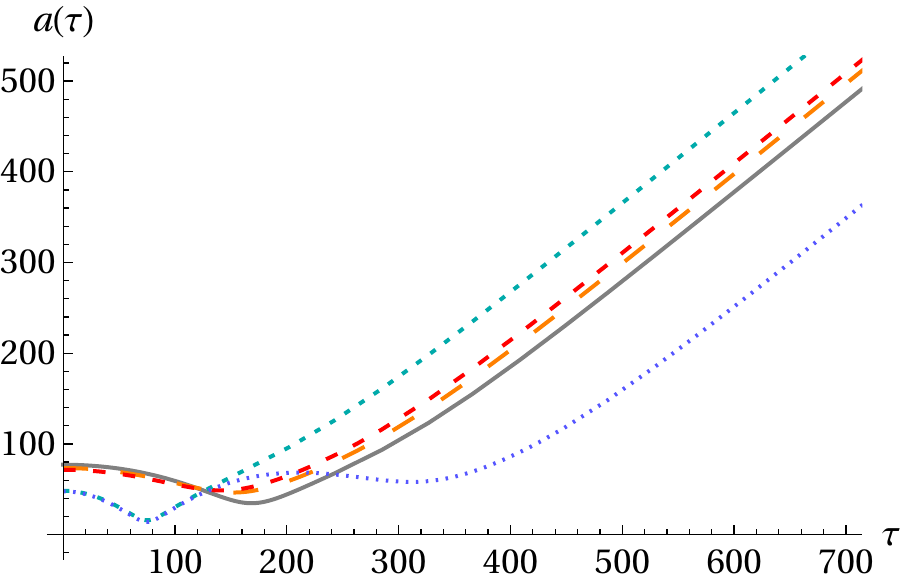}
	\includegraphics[width=0.45\textwidth]{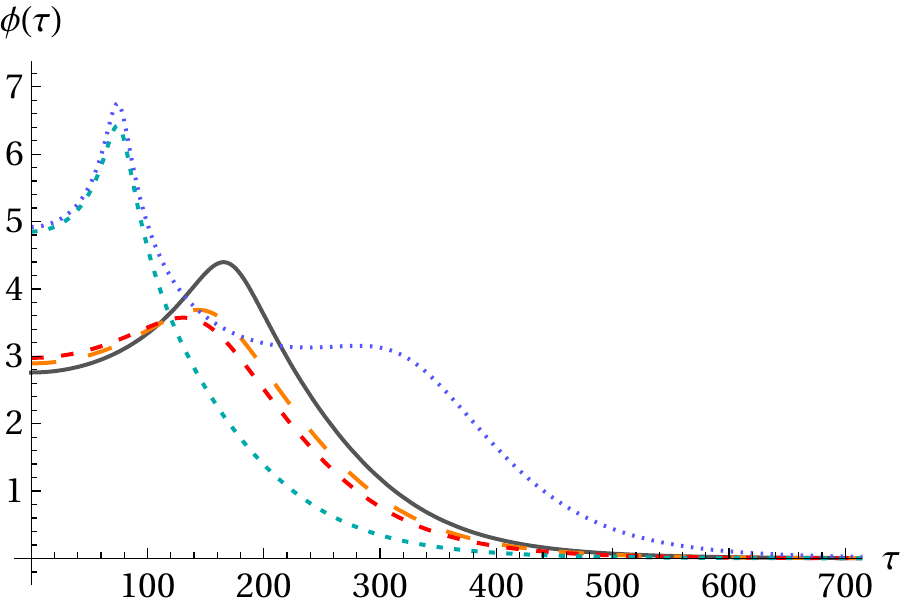}
	\caption{Expanding wormholes with dilaton coupling $\beta=1.4$ and $m^2 N=3.525.$ The initial dilaton values are $\phi_0=2.7593083935,$ $2.8947797102,$ $2.9770027664,$ $4.8514743456,$ $4.9204400124.$}\label{fig:GS14exp}
\end{figure}

\begin{figure}
	\includegraphics[width=0.45\textwidth]{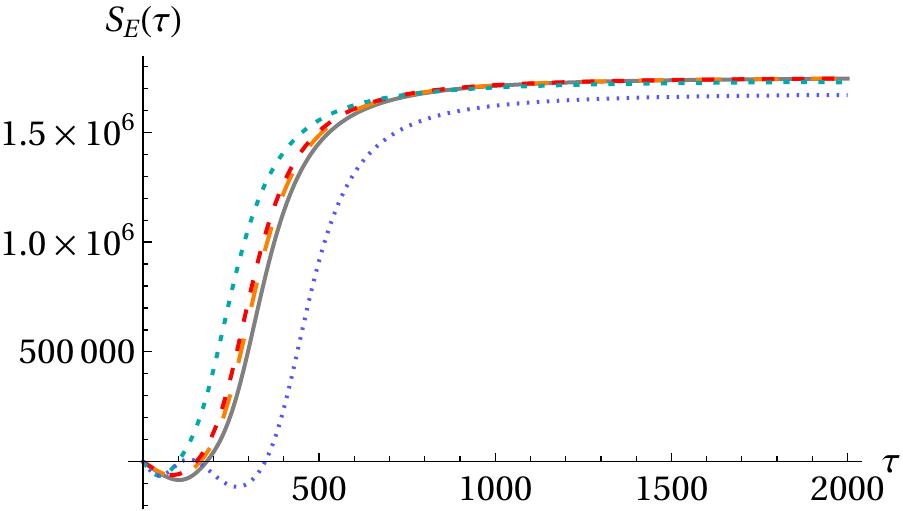}
	\includegraphics[width=0.45\textwidth]{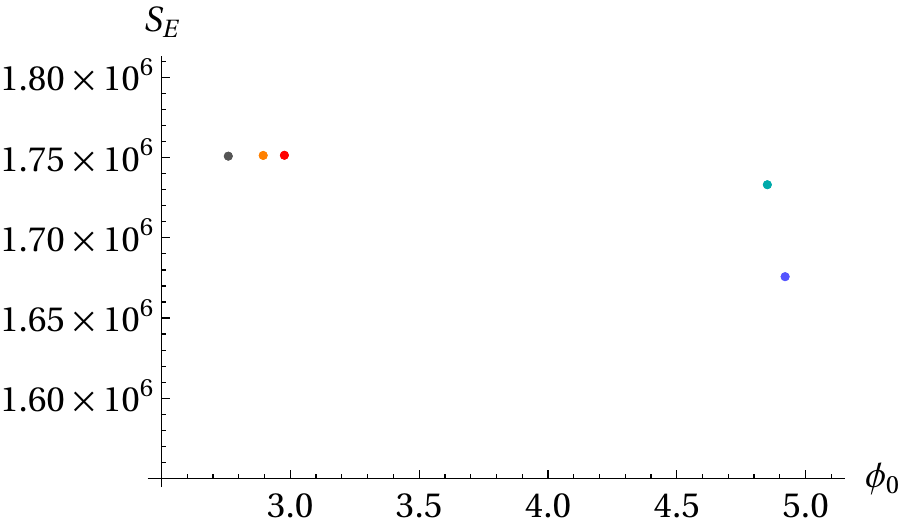}
	\caption{The Euclidean action of the solutions shown in Fig.~\ref{fig:GS14exp}. Once again, the solutions with additional oscillations have lower Euclidean action.}\label{fig:GS14expS}
\end{figure}

The action for these expanding wormholes is shown in Fig.~\ref{fig:GS12expS}, on the left as a function of radius and on the right in terms of the asymptotic values. This time, in contrast with Fig.~\ref{fig:GS12S}, the action immediately starts decreasing as the solutions progressively develop additional inflection points and oscillations. This unusual feature complicates the interpretation of these solutions, as it naively suggests that expanding wormholes with more oscillations are more likely than those with fewer features. We will comment further on this peculiarity in the discussion section. Similar features occur at other values of the dilaton coupling, see Figs.~\ref{fig:GS14exp} and \ref{fig:GS14expS} for additional examples.

\begin{figure}
	\includegraphics{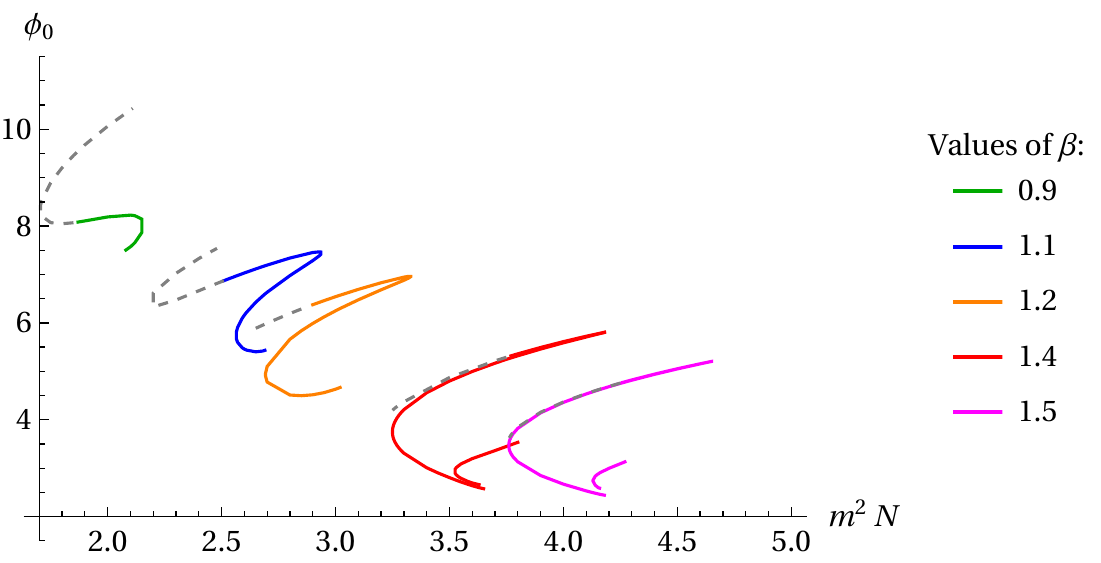}
	\caption{Summary plot of the existence of expanding wormholes, as a function of the mass/charge combination $m^2 N$ and the initial dilaton values $\phi_0,$ for various values of the dilaton coupling $\beta.$ A detailed description is provided in the main text.} \label{fig:expworm}
\end{figure}

We can now try to gain a more global understanding of the existence of expanding wormhole solutions. The existence of solutions as a function of $m^2 N$ and $\phi_0,$ for various values of the coupling $\beta,$ is shown in Fig.~\ref{fig:expworm}. The first impression that one obtains is that, unlike for generalised GS-type wormholes, the behaviour is qualitatively similar for different $\beta$ values. In this case, no solutions exist at zero mass, hence we do not expect (and do not see) branches connecting to the vertical axis. However, at sufficiently large $m^2 N$ branches of solutions develop, starting at values at which a gap in the effective potential $W(\phi)$ appears, in analogy with the GS-type solutions and as discussed in connection with Fig.~\ref{fig:W}. These branches then continue in a rough zig-zag fashion towards larger $\phi_0$ values. As they do so, the solutions become progressively more intricate, developing inflection points and then additional oscillations. The dashed lines indicate parameter regions where an additional minimum in the scale factor exists, and one should picture the branches as continuing upwards indefinitely, with the solutions containing ever greater numbers of oscillations.

\begin{figure}[h!]
	\includegraphics[width=0.31\textwidth]{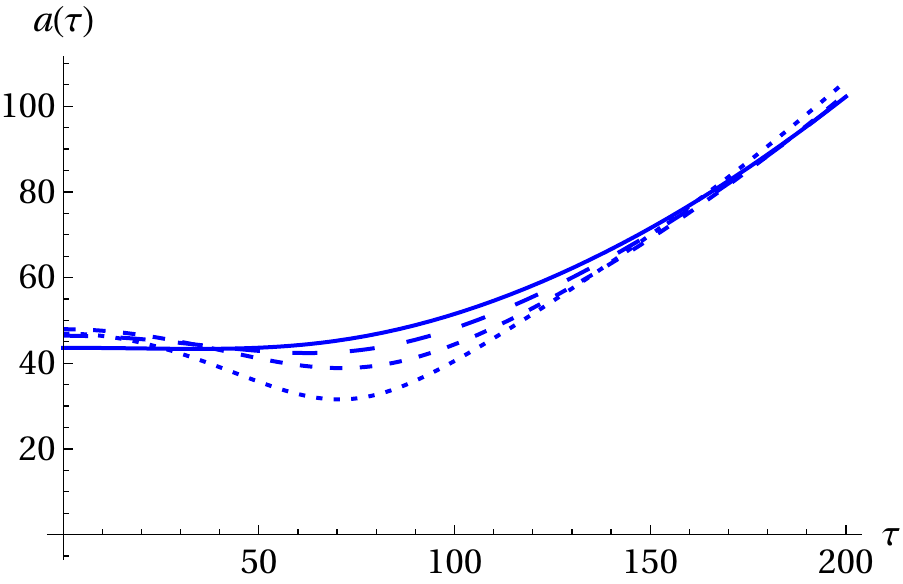}
	\includegraphics[width=0.31\textwidth]{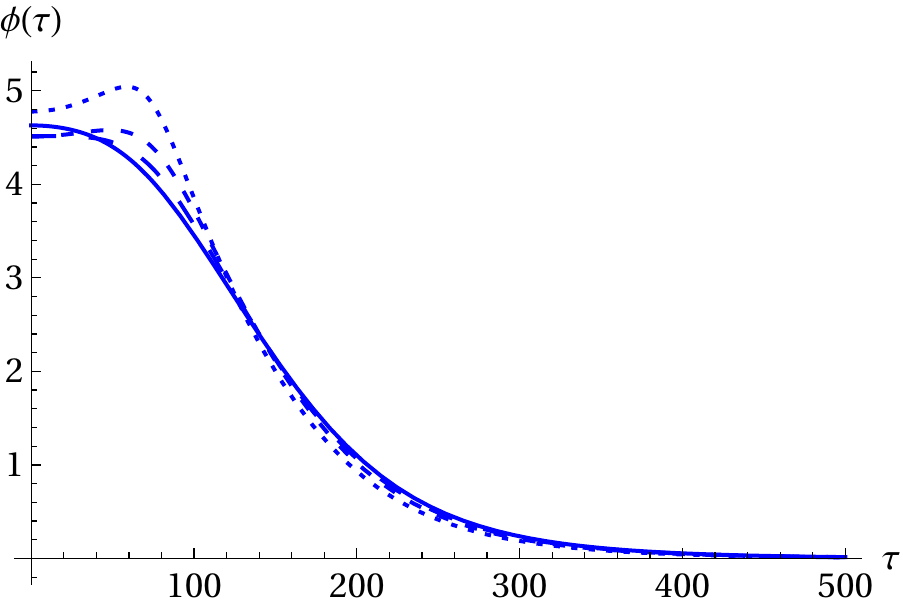}
	\includegraphics[width=0.31\textwidth]{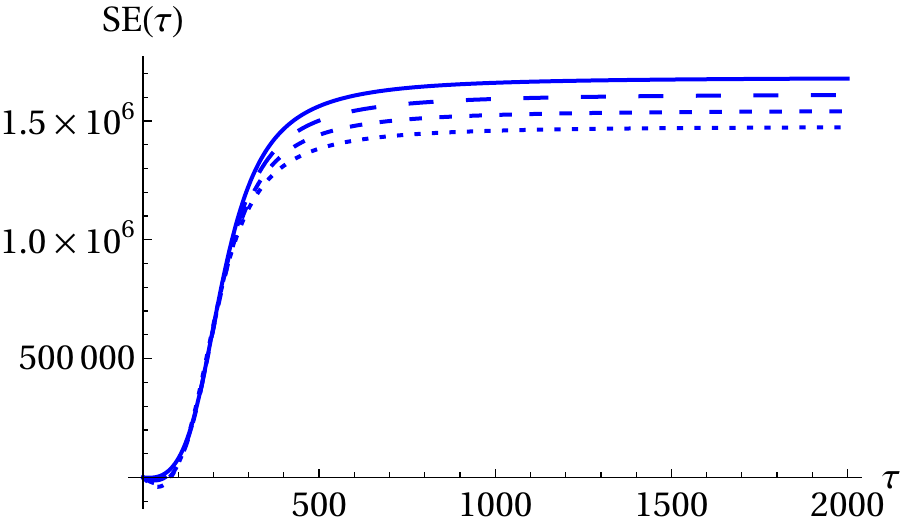}\caption{Transition from a solution with a monotonically evolving dilaton to a solution with an oscillating dilaton, with parameter values $\beta=1.2$ and $N=30000$ to $27000$, the smaller values of $N$ corresponding to the more finely dashed curves.} \label{fig:inflpoint}
\end{figure}

\begin{figure}[h]
	\includegraphics[width=0.31\textwidth]{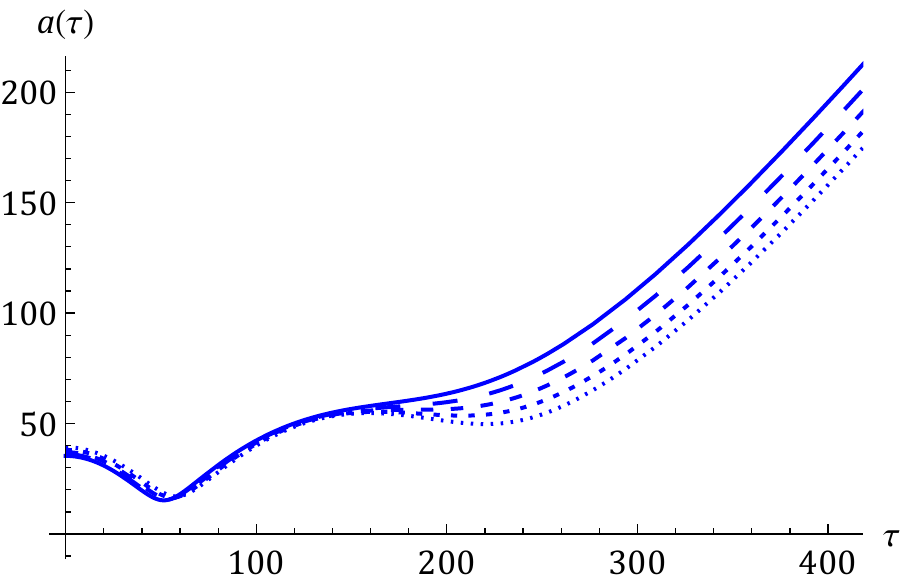}
	\includegraphics[width=0.31\textwidth]{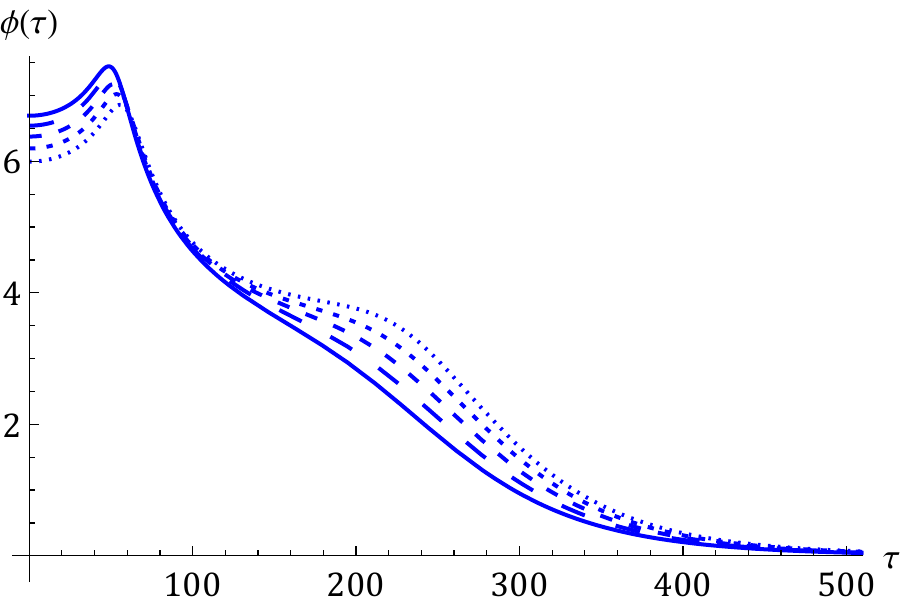}
	\includegraphics[width=0.31\textwidth]{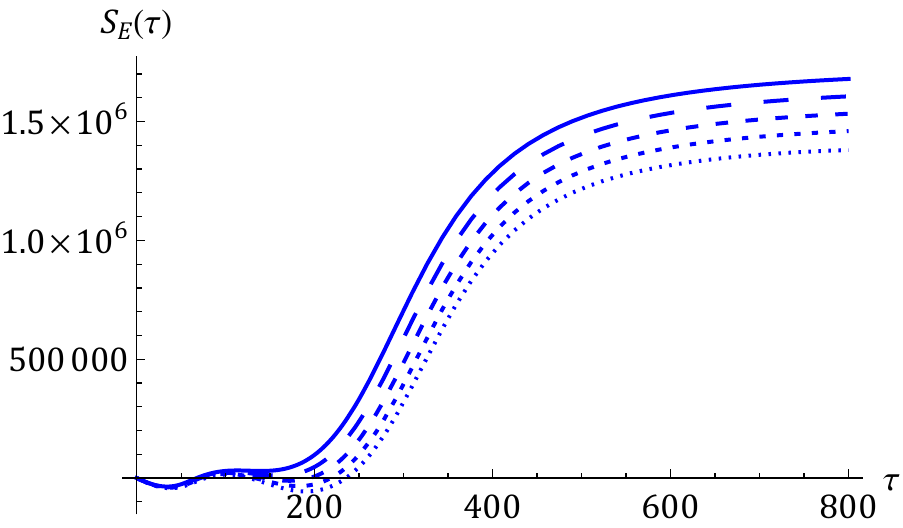}
	\caption{Transition from a solution with a single minimum of the scale factor to one with two minima. The parameter values are $m=0.01, \beta=1.2$ and  $N=31000, \phi_0=6.6921845703; N=30000, \phi_0 = 6.5411318359; N=29000, \phi_0 = 6.3756220703; N=28000, \phi_0 = 6.1942070312; N=27000,\phi_0=5.9951564438$ (smaller values of $N$ correspond to more finely dashed curves).} \label{fig:secondmin}
\end{figure}

Two examples of series of solutions, which show how additional features form, are provided in Figs.~\ref{fig:inflpoint} and \ref{fig:secondmin}. In the first case, we see how the dilaton starts changing its initial direction of evolution. In the second case, we see how an additional minimum in the scale factor develops. In both series, it is again made manifest that additional features lower the action.

\begin{figure}[h!]
	\includegraphics[width=0.31\textwidth]{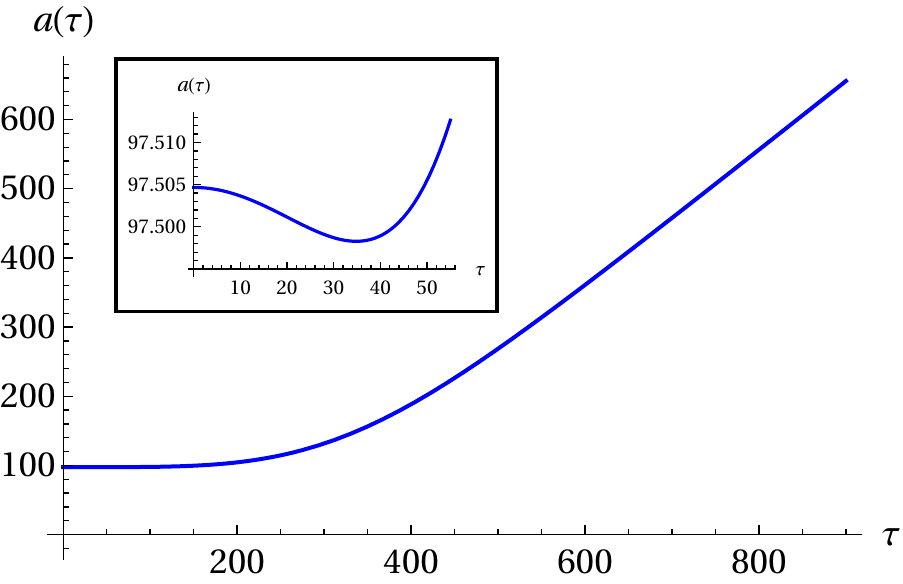}
	\includegraphics[width=0.31\textwidth]{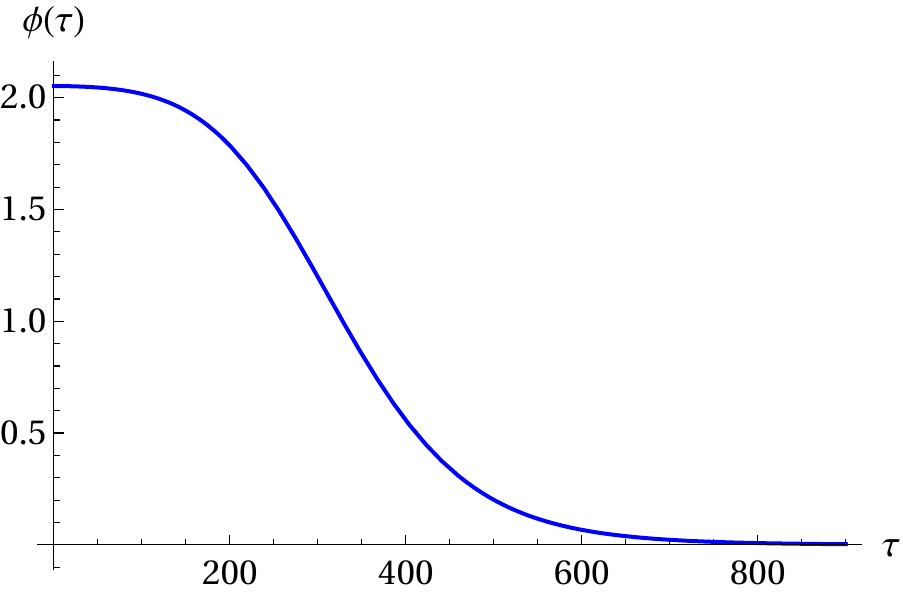}
	\includegraphics[width=0.31\textwidth]{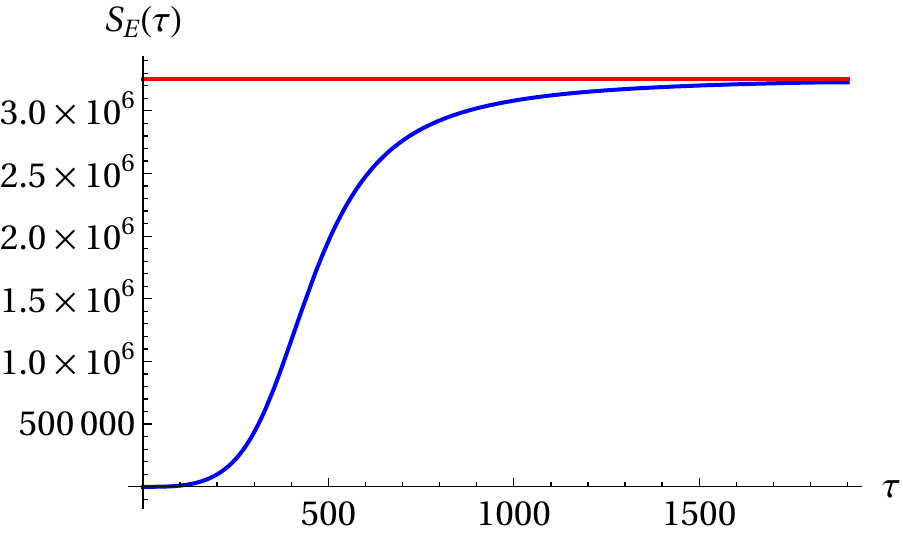}
	\caption{An example of a wormhole with large dilaton coupling, in this case $\beta=2.$  Shown are again the scale factor (left), dilaton (middle) and Euclidean action (right). The parameter values are $m=0.01,$ $\beta = 2,$ $N=73940$ and the initial dilaton value is $\phi_0=2.0522333714.$} \label{fig:betatwo}
\end{figure}

One feature which we have not mentioned yet is that for large enough dilaton coupling (approximately $\beta>1.3$) a second branch of solutions forms, see the examples with $\beta=1.4, 1.5$ in Fig.~\ref{fig:expworm}. These additional branches are again bifurcating, ending on both sides at the locations where the effective potential stops developing a gap. The expanding wormhole solutions belonging to these branches are however very similar to those in the first branch. In fact, we already saw examples of solutions belonging to a second branch in Figs.~\ref{fig:GS14exp} and \ref{fig:GS14expS} -- these were the solutions with $\phi_0 \approx 2.89, 2.97$. These figures indicate that the various solutions evolve continuously with the initial value $\phi_0,$ and that no significant physical distinction between the branches is noticeable. We should also highlight that there appears to exist no upper bound on the dilaton coupling $\beta.$ Although solutions do become harder to find numerically at large $\beta,$ we see no evidence for any obstruction at large $\beta$ -- for illustration, we provide an example of an expanding wormhole with $\beta=2$ in Fig.~\ref{fig:betatwo}. The existence of solutions at large dilaton coupling may be of interest in string theory models, where it is rather natural to obtain dilaton couplings of order unity.

\begin{figure}
	\includegraphics[width=0.6\textwidth]{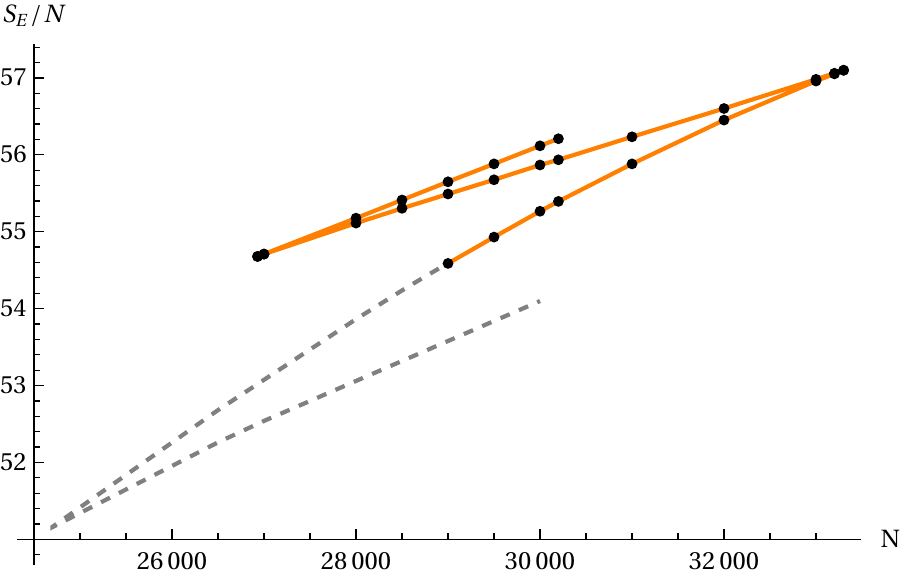}\caption{The action-to-charge ratio for expanding wormholes with $\beta=1.2.$ The dashed line represents solutions that contain a second minimum in the scale factor, {\it cf.} also the $\beta=1.2$ locus in Fig.~\ref{fig:expworm}. In all cases, the action-to-charge ratio increases with increasing charge.} \label{fig:actcharge}
\end{figure}

Fig.~\ref{fig:actcharge} explores a further physical aspect of the solutions, namely their action-to-charge ratio. As the figure indicates, this is found to be a monotonically increasing function of the charge $N$ (with fixed mass $m$), which points to the fact that the expanding wormhole solutions might be non-perturbatively unstable, as it would be preferable for a fixed charge wormhole to break up into two smaller wormholes with combined charge equal to the original one \cite{Andriolo:2022rxc}, as long as appropriate solutions with smaller $m^2 N$ actually exist (Fig.~\ref{fig:expworm} indicates that if they exist, they must contain a large number of oscillations). We should note however that determining the rate of this process would require knowledge of interpolating solutions linking the initial and final configurations. We leave this topic for future research.

\begin{figure}
	\includegraphics[width=0.3\textwidth]{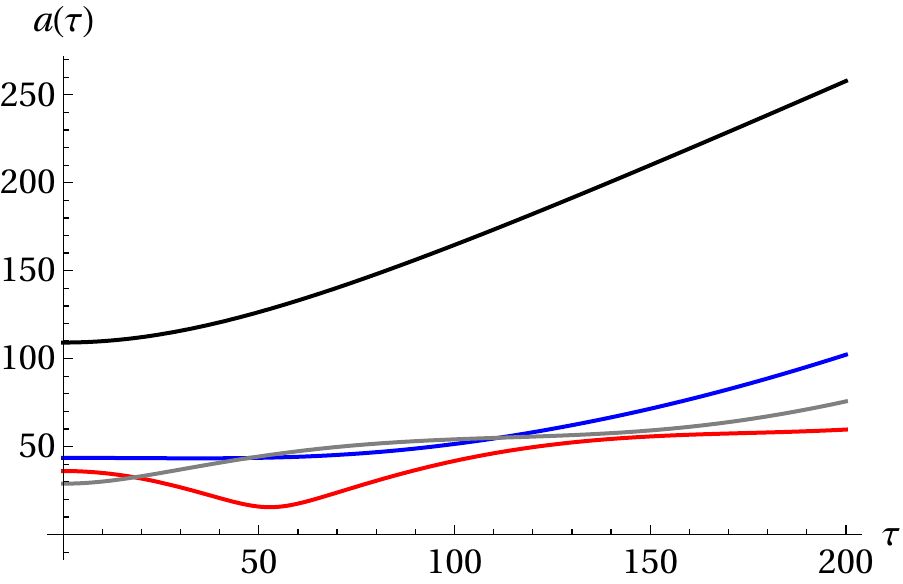}
	\includegraphics[width=0.3\textwidth]{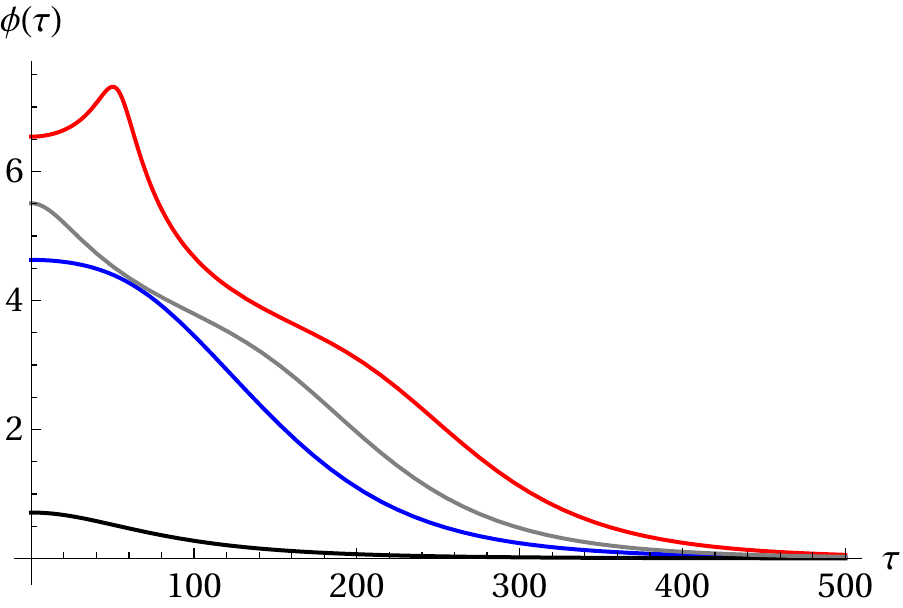}
	\includegraphics[width=0.3\textwidth]{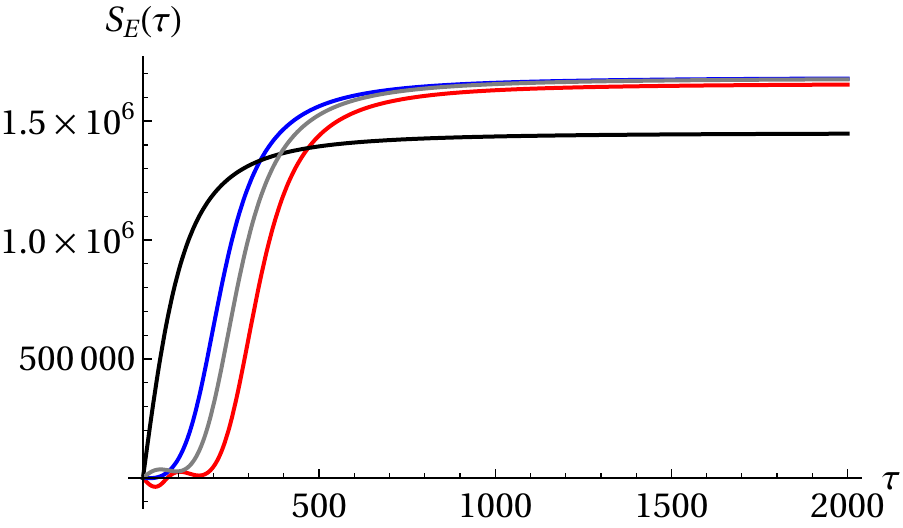}
	\caption{This figure compares GS-type and expanding wormholes, at the same parameter values $m^2 N = 3$ and dilaton coupling $\beta=1.2.$ The GS-type solutions are depicted by the black ($\phi_0 \approx 0.7$) and grey ($\phi_0 \approx 5.5$) lines. These solutions were already presented in Fig.~\ref{fig:GS12}. The blue ($\phi_0 \approx 4.6$) and red ($\phi_0 \approx 6.2$) curves correspond to expanding wormholes, and were shown in Fig.~\ref{fig:GS12exp}. Interestingly, the actions are seen to be quite close to each other, with the grey solution lying in between the two expanding wormhole solutions. It appears that overall the black GS-type solution is dominant, but a verification of this assertion would require an understanding of the infinite oscillation limit of expanding wormholes. }\label{fig:GScomparison}
\end{figure}

Finally, we may compare GS-type and expanding wormholes, at fixed parameter values. This is done in Fig.~\ref{fig:GScomparison}. As one can see in this picture, the different types of solutions exist at nearby initial values of the dilaton, and in fact even the values of the Euclidean action are intertwined. Overall, the simplest solution, {\it i.e.} the GS-type solution with monotonically evolving dilaton, appears to have the lowest action. However, as we cannot extrapolate our results to large numbers of oscillations in the fields, it remains an open question whether this solution is truly the most dominant one. At larger values of the dilaton coupling, we already know that the simplest solution is not the one with the lowest action, {\it cf.} Fig.~\ref{fig:GS158S}. This is an interesting puzzle requiring future work, and we will provide some comments related to this question in the discussion section.

\section{Axion-scalar wormholes leading to expanding baby universes} \label{sec:axionscalar}

\begin{figure}[h]
	\includegraphics[width=8cm]{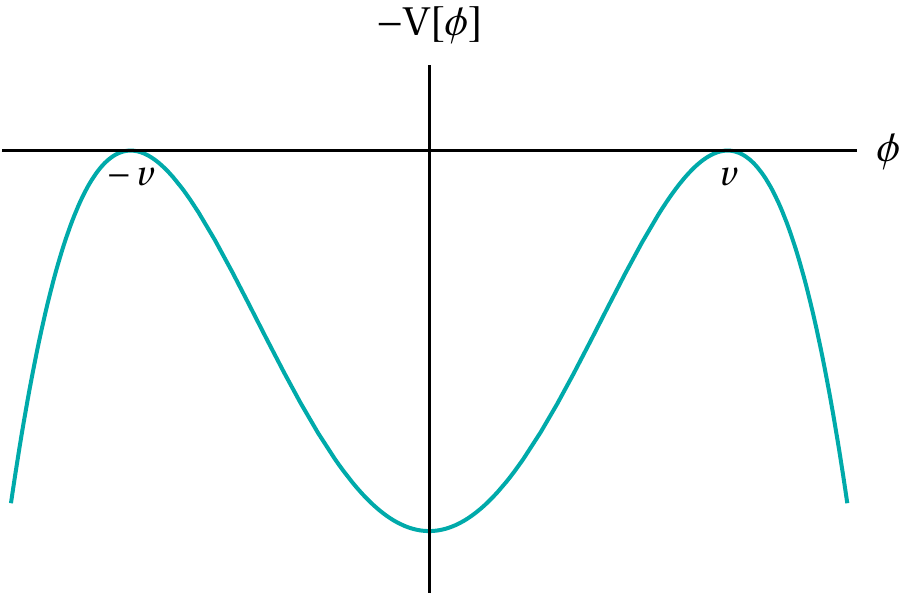}\caption{Symmetric double well potential for the parameter values $\lambda=0.01$ and $v=0.4$.}\label{fig:symmetricpotential}
\end{figure}

In the previous section, we saw that the dilaton coupling had a significant effect on the existence and properties of wormhole solutions. One might wonder if solutions also exist without the special dilatonic coupling, {\it i.e.} in the presence of an ordinary scalar field. Thus we will set $\beta=0,$ but we will keep the letter $\phi$ to designate the scalar. The equations of motion \eqref{eq:fulleomh1} then read
\begin{equation}
	\left\lbrace
	\begin{aligned}
		&2a\ddot{a}+\dot{a}^2-1+\kappa a^2\left(\frac{\dot{\phi}^2}{2}+V(\phi)\right)-\frac{\kappa N^2}{a^4}=0\quad\text{(acceleration equation)}\,,\\
		&\dot{a}^2-1=\frac{\kappa a^2}{3}\left(\frac{\dot{\phi}^2}{2}-V(\phi)\right)-\frac{\kappa N^2}{3 a^4}\quad\text{(Friedmann constraint)}\,,\\
		&\ddot{\phi}+\frac{3\dot{a}}{a}\dot{\phi}=\frac{\dd V}{\dd\phi}\quad\text{(scalar equation)}\,.
	\end{aligned}
	\right.\label{eq:scalarfieldeom}
\end{equation}
We will choose the scalar field potential to be of double well form
\be
V(\phi) = \frac{1}{4} \lambda (\phi^2 - v^2)^2 \kma \label{eq:symmetricV}
\ee
where $\lambda$ is a dimensionless scalar field self-coupling and $v$ is the vacuum expectation value, see Fig.~\ref{fig:symmetricpotential}. The self-coupling $\lambda$ may be scaled to any convenient value using the rescalings detailed in appendix \ref{sec:scaling}.

The scalar field equation in \eqref{eq:scalarfieldeom} possesses a simple mechanical analogy as the motion of a ``particle'' in an inverted potential $-V$, under the velocity dependent friction force,
\be
F_\text{friction}=\frac{3\dot{a}}{a}\dot{\phi} \pkt
\ee
We see that the friction coefficient could be positive or negative depending on the sign of $\dot{a}$.
Since we are looking for asymptotically flat solutions,
the scalar field should approach one of the vacua  as $\tau \to \infty$. Without loss of generality we choose this to be the left vacuum, $\phi=-v$.
A particle released with zero velocity will get a chance to reach this vacuum only if it starts at some $\phi_0 \in [0, v]$.
Put differently, the potential barrier is required in order to make the scalar field roll in the appropriate direction after starting with zero velocity. Moreover, the potential minimum is required in order to stabilise the field asymptotically and to approach flat spacetime. For this purpose a single minimum would be enough, but for simplicity (scarcity of free parameters) we choose a symmetric potential here.
Typically if $\phi_0$ is very close to $v$ we get ``over-shooting'', since for small $\tau$ the scale factor derivative $\dot{a}<0$  and we have anti-friction.
Generally with the simple integration of equation \eqref{eq:scalarfieldeom} we can conclude that the
work of the friction force has to be compensated by the potential energy \cite{Lavrelashvili:1987zx,Rubakov:1988wx},
\be
3 \int_{0}^{\infty} d\tau \frac{3\dot{a}}{a}{\dot{\phi}}^2 = - V(\phi_0) \pkt \label{eq:balance1}
\ee
The condition $\Delta>0$ implies for our symmetric double well potential that:
\be
	4-\kappa^3N^2\frac{\lambda^2}{16}(\phi_0^2-v^2)^4>0\quad \Leftrightarrow\quad (\phi_0^2-v^2)^4<\frac{64}{\kappa^3N^2\lambda^2}
	\Leftrightarrow\quad  v^2-\phi_0^2<\frac{2\sqrt{2}}{\kappa^{3/4}\sqrt{N\lambda}} \pkt
\ee
For $\phi_0 \in [0, v]$, this means that we are restricted to a range of values:
\begin{equation}
	\sqrt{v^2-\frac{2\sqrt{2}}{\kappa^{3/4}\sqrt{N\lambda}}}<\phi_0 < v \pkt \label{eq:phi0}
\end{equation}
For each set of parameters $(v,N)$, a wormhole solution is found by fine-tuning the value of $\phi_0$ between under-shooting and over-shooting values.
This implies that, if for the minimal value allowed by the condition \eqref{eq:phi0},
\be
\phi_0^\text{min}=\sqrt{v^2-\sqrt{8/(\kappa^{3/2}N\lambda)}} \kma
\ee
we have over-shooting, then there will not exist any wormhole solutions.
Conversely, if $\phi_0=\phi_0^\text{min}$ leads to under-shooting, then there will necessarily exist a wormhole solution.
Therefore, for each value of $v$, the above criterion defines an upper bound on the value of $N$ for which a wormhole solution exists.

Note also that in the symmetric potential Eq.~(\ref{eq:symmetricV}) no GS-type wormholes exist with
{\it non-trivial} scalar field. This conclusion can be easily reached by noting that GS-type wormholes have
positive $\dot{a}$, and consequently provide friction in the scalar field equation.
So, in order for the solution to end up on one of the hills (vacua) for $\tau \to \infty$
one should start rolling down from higher hill, in order to overcome friction,
but in a symmetric potential there is no higher hill. We conclude that the only known GS wormholes 
in this potential arise with a trivial scalar field configuration $\phi = \pm v$.

\begin{figure}[h]
	\includegraphics[width=0.31\textwidth]{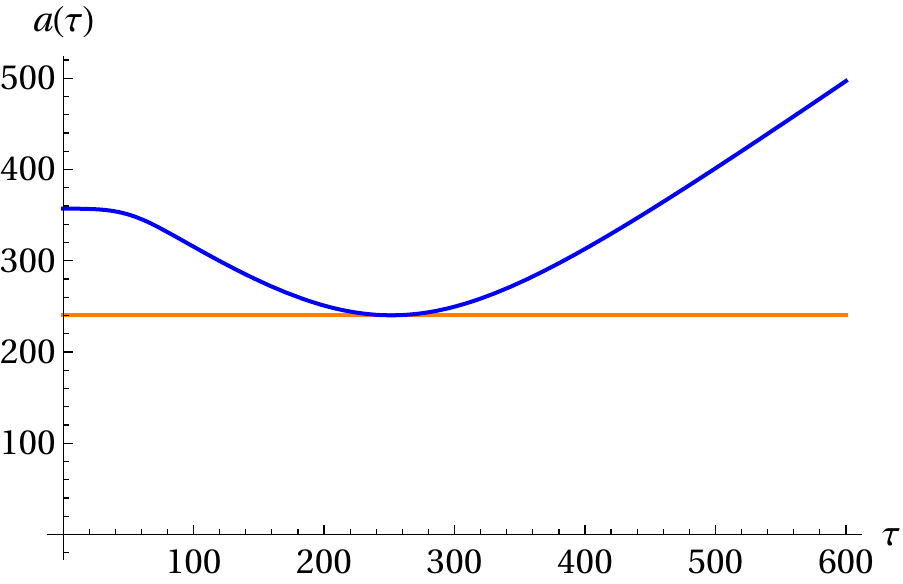}
	\includegraphics[width=0.31\textwidth]{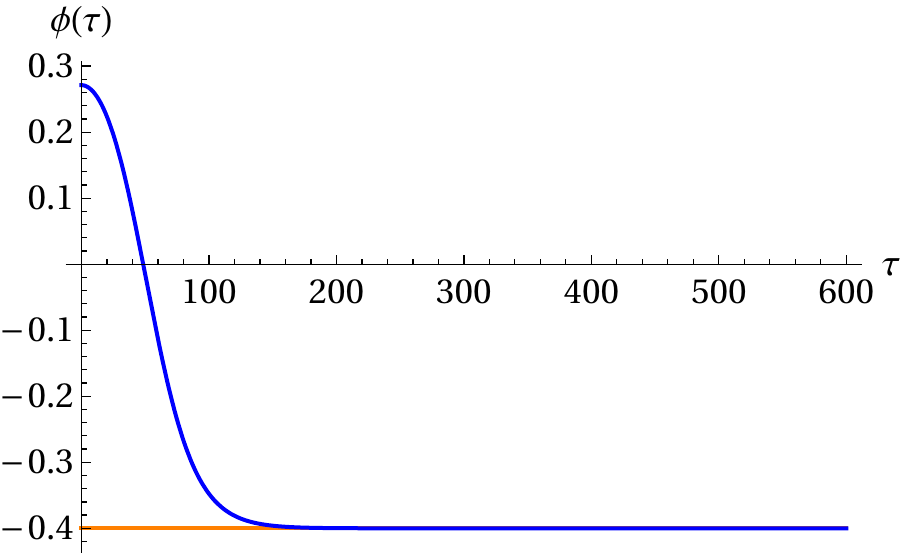}
	\includegraphics[width=0.31\textwidth]{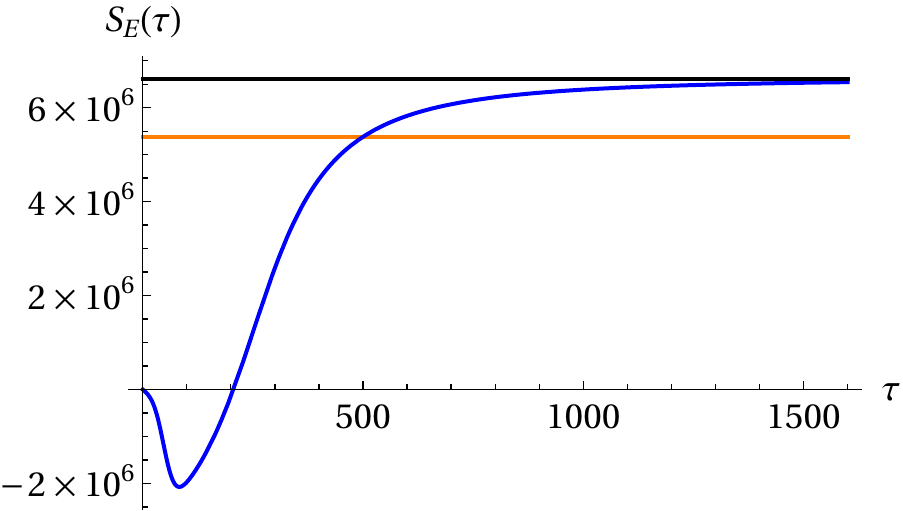}
	\caption{An example of an expanding wormhole supported by a large axionic charge, and a scalar field in a double well potential. The parameters used are  $\lambda=0.01\,,\ N=100000\,,\ v=0.4$. The initial scalar field value is $\phi_0=0.27112946714882599307.$ The orange lines provide the GS wormhole values as reference.} \label{fig:examplelarge}
\end{figure}

\begin{figure}[h]
	\includegraphics[width=0.31\textwidth]{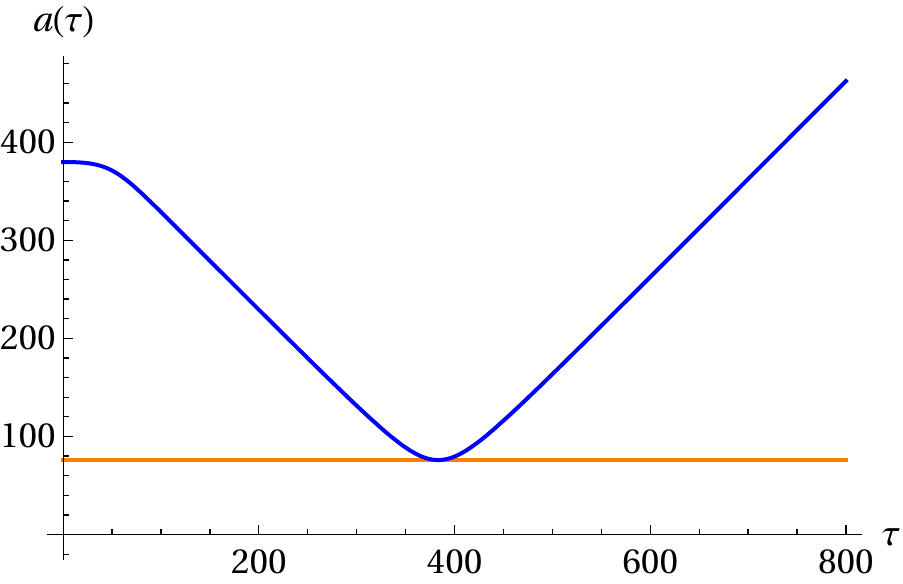}
	\includegraphics[width=0.31\textwidth]{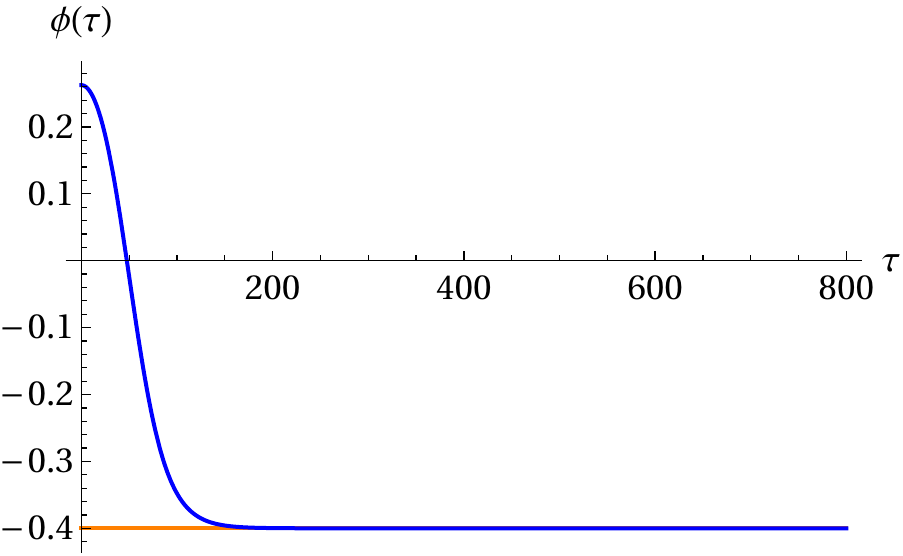}
	\includegraphics[width=0.31\textwidth]{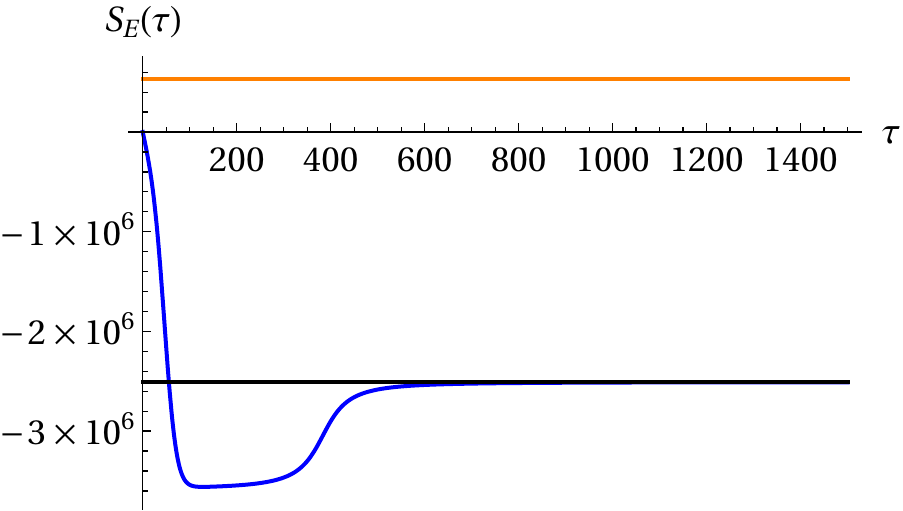}
	\caption{An example of an expanding wormhole supported by a small axionic charge. The parameters used are  $\lambda=0.01\,,\ N=1000\,,\ v=0.4$. The initial scalar field values is $\phi_0=0.26235021388116072967.$ } \label{fig:examplesmall}
\end{figure}

\begin{figure}[h]
	\includegraphics[width=0.5\textwidth]{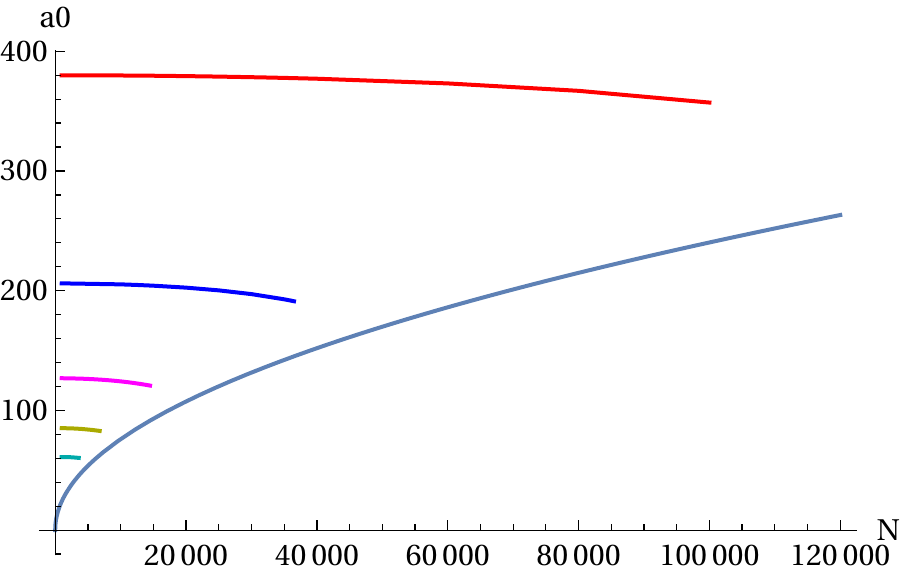}
\caption{Wormhole throat size $a_0$ vs $N$ for the parameter values $\lambda=0.01$ and $v=0.4, 0.5, 0.6, 0.7, 0.8$ (going down in the plot).
At large $N$ all curves tend to the GS value, shown here by the blue curve.}\label{fig:a0againstN}
\end{figure}

For the reasons just given we will rather look for wormholes that have a local maximum at the throat, that is to say wormholes that lead to expanding baby universes after analytic continuation to Lorentzian time. The first examples are shown in Figs.~\ref{fig:examplelarge} and \ref{fig:examplesmall}, for a large and a small axionic charge respectively. Compared to the dilatonic case, we see that the local maximum at the origin is much more pronounced, and even more so for the small charge solution. The minimum value of the scale factor is found to coincide rather precisely with the GS value (by which we mean the throat size of a GS wormhole, with the scalar residing at a potential minimum), indicated by the orange lines in the figures. In fact, as the charge grows, the initial throat size is reduced, and one may infer that the solutions with larger charge approach GS-type solutions more and more. This is confirmed by a study of the initial throat size as a function of the charge, shown for various potential widths in Fig.~\ref{fig:a0againstN}.

The Euclidean action for these solutions is shown in the right panels of Figs.~\ref{fig:examplelarge} and \ref{fig:examplesmall}, as a function of the radial coordinate. One thing we may notice is that it starts off by obtaining negative contributions near the throat. Further away from the throat, as the scalar potential becomes less important (since the scalar approaches the vacuum), the action receives positive contributions. Which contributions dominate depends rather crucially on the axionic charge: the large charge solution ends up having a positive action (in this case larger than the GS value), while the small charge wormhole has a negative Euclidean action. As we will discuss further below, this suggests that only large charge wormholes might be of physical relevance.

\begin{figure}
	\begin{subfigure}{0.49\linewidth}
		\includegraphics[width=0.8\textwidth]{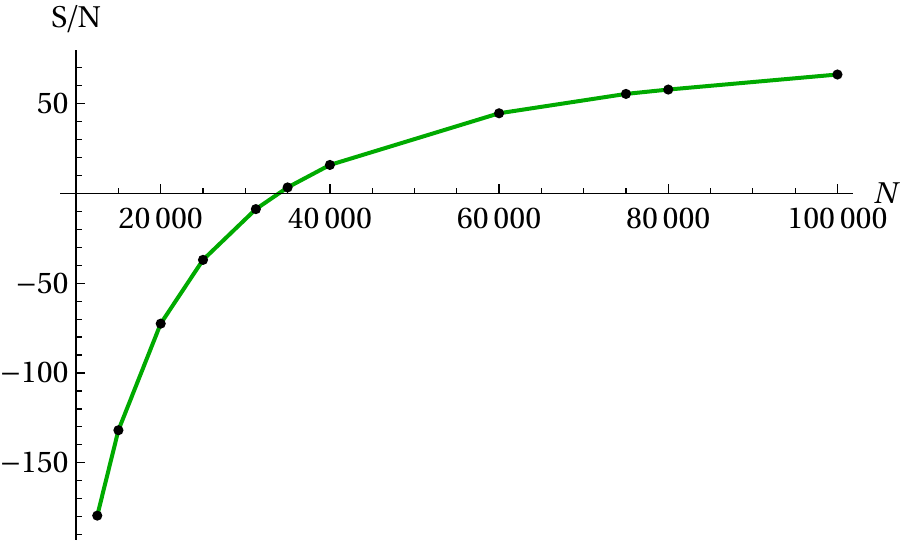}\subcaption{$v=0.4$}
	\end{subfigure}
	\begin{subfigure}{0.49\linewidth}
		\includegraphics[width=0.8\textwidth]{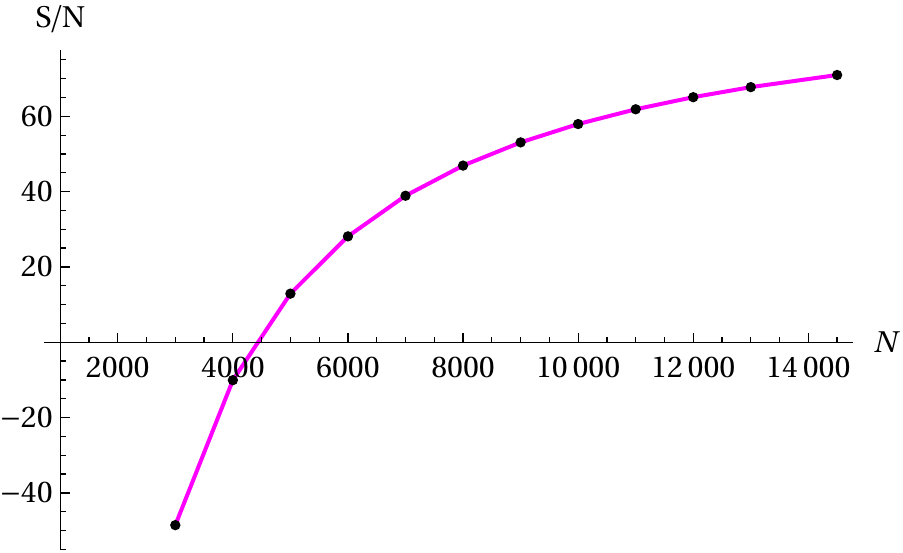}\subcaption{$v=0.6$}
	\end{subfigure}
	\caption{Plots of the action-to-charge ratio for $v=0.4$ (left) and $v=0.6$ (right) with $\lambda = 0.01$.} \label{fig:atc}
\end{figure}

Fig.~\ref{fig:atc} shows the action-to-charge ratio for two different potential widths. Just as for the dilatonic case, this ratio is a monotonically increasing function of the charge, indicating that such wormholes could break up non-perturbatively into smaller components.

\begin{figure}
	\includegraphics[width=0.31\textwidth]{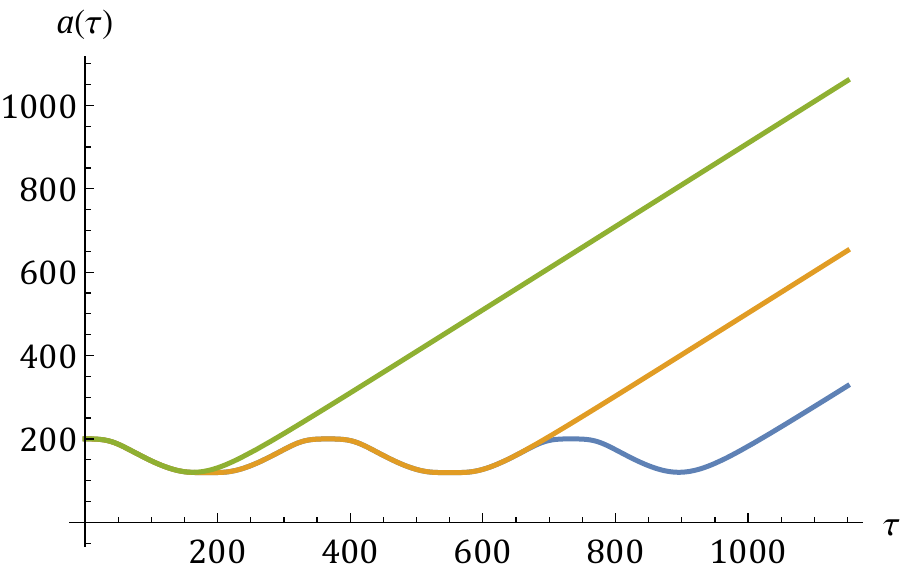}
	\includegraphics[width=0.31\textwidth]{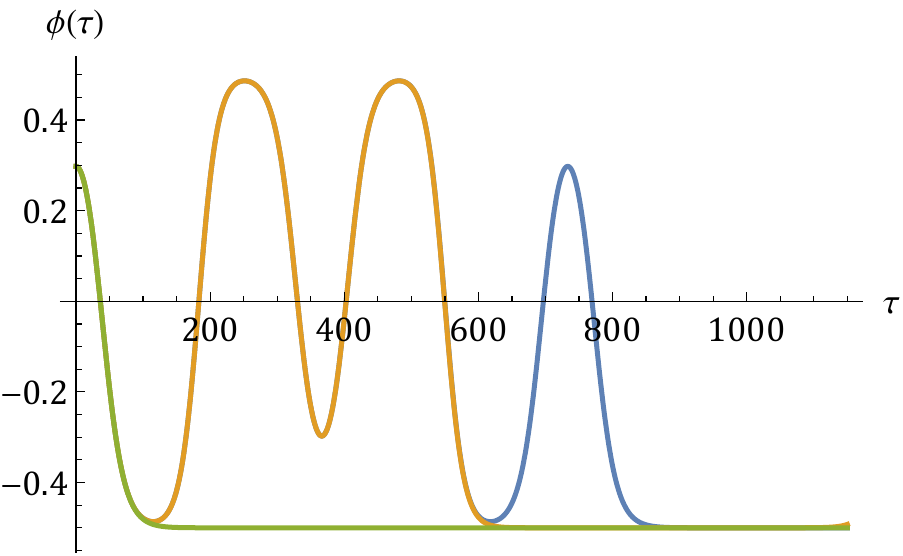}
	\includegraphics[width=0.31\textwidth]{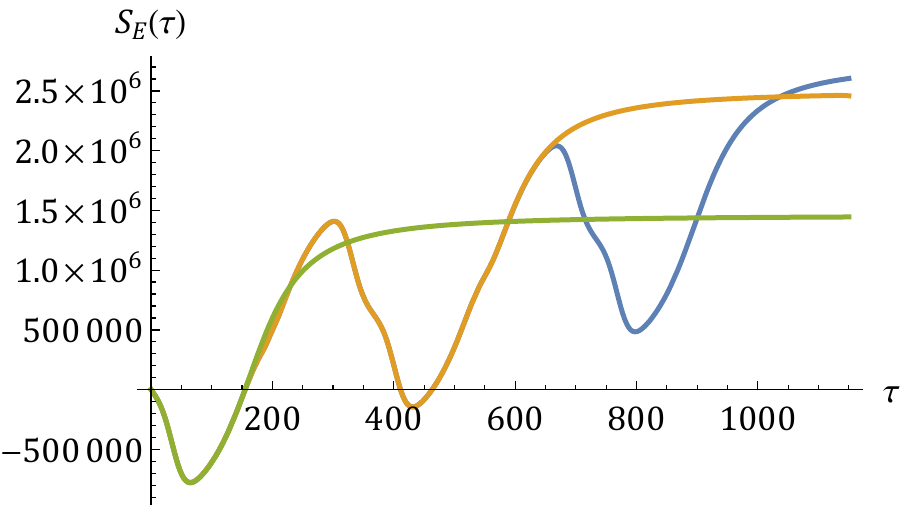}\caption{Comparison of solutions with one, two and three minima for the same theory parameters $N=25000$, $\lambda=0.01$ and $v=0.5$. The Euclidean action grows with each additional oscillation. The solutions here are specified by initial scalar field values that lie very close to each other, respectively at $\phi_0=0.297695980172969317414540,$ $ 0.297530409785421517546558,$ $0.297530409646648251937091$ (these solutions must be optimised to high accuracy in order to determine the action reliably).} \label{fig:oscillations}
\end{figure}

In this theory there also exist solutions with additional oscillations in the fields, a property first observed in  \cite{Lavrelashvili:1996tk}, see the example in Fig.~\ref{fig:oscillations}. As one can see there, each oscillation simply adds to the previous solution. That is to say, the solutions with additional oscillations do not display completely different field evolutions, but rather seem to be extensions of solutions with fewer oscillations. As one can also clearly see in the right panel, each oscillation adds a positive contribution to the action, rendering it larger and larger. This implies that oscillating solutions are associated with a lower probability, which is physically reasonable (and differs from the dilatonic case).

\begin{figure}
	\includegraphics[width=0.45\textwidth]{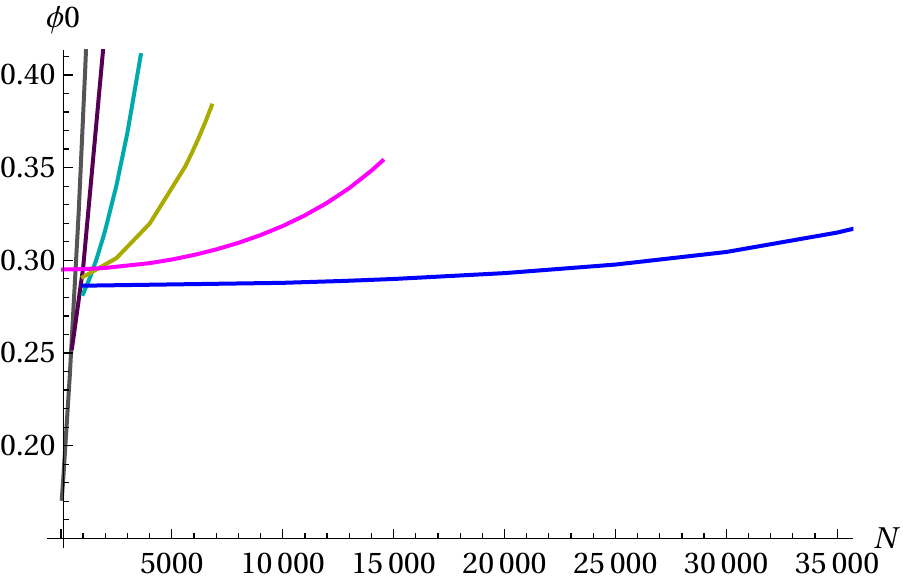}
	\includegraphics[width=0.45\textwidth]{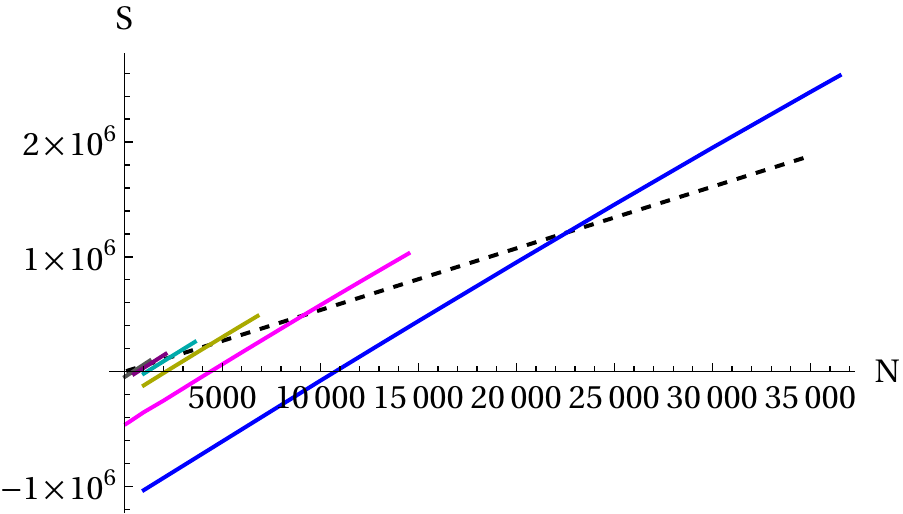}
	\caption{Summary plots for $v=1.0$ (black), $v=0.9$ (purple), $v=0.8$ (turquoise), $v=0.7$ (green), $v=0.6$ (pink) and $v=0.5$ (blue). Here $\lambda=0.01.$ The right plot shows the most striking result, namely that the action varies linearly with the charge $N.$ For small enough charge, the action becomes negative. The dashed black line is the Giddings-Strominger value of the action.} \label{fig:scalar}
\end{figure}

In this theory it turns out that the existence of solutions as a function of the axionic charge $N$ is simpler than in the dilatonic case, see Fig.~\ref{fig:scalar}. The initial $\phi_0$ values lie on monotonic curves. However, the Euclidean action has more notable features, see the right panel in the figure. What jumps out most clearly is that the Euclidean action varies essentially linearly with the charge $N,$ though it is not proportional to it. Also, the action is larger than the GS value at sufficiently large $N,$ but decreases significantly at small $N$ and even reaches negative values, as noticed before in the example of Fig.~\ref{fig:examplesmall}.

\begin{figure}
	\includegraphics[width=0.6\textwidth]{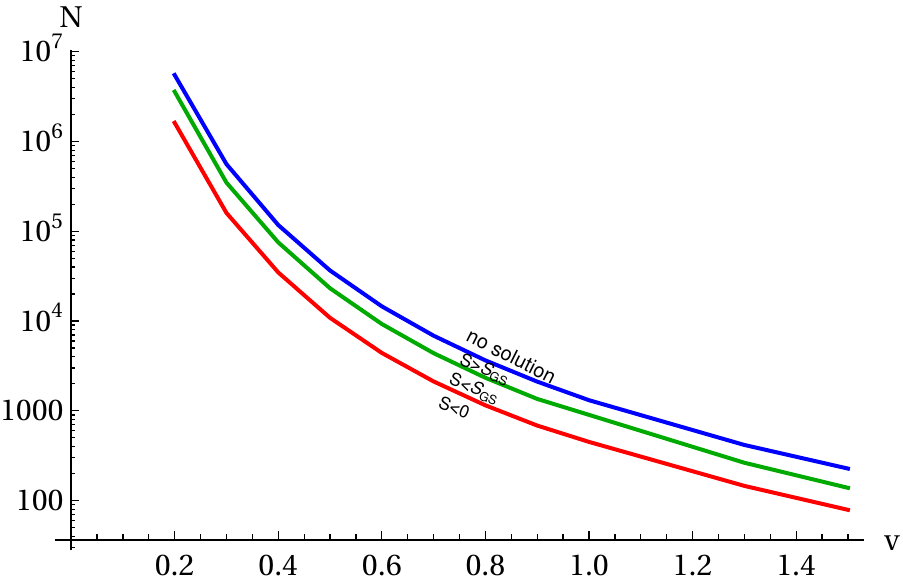}
	\caption{Existence of expanding wormhole solutions in a double well potential with width $v,$ and for axionic charge $N$. Results for $v \in [0.2,1.5]$ have been incorporated. Here $\lambda = 0.01$.}\label{fig:summarydoublewell}
\end{figure}

The overall structure of the solutions is summarised from a different point of view in Fig.~\ref{fig:summarydoublewell}, as a function of the charge and the potential width. Solutions exist below the blue line, have action smaller than GS below the green line, and negative action below the red line. Thus, although solutions exist over vast regions of parameter space, it is only in the narrow band between the red and blue lines that we expect to find physically relevant solutions. Especially in the even narrower band between the red and green lines, we may even expect these solutions to play an important role, as they have higher probabilities associated to them than the corresponding GS wormholes.

\section{Conclusions}

We have investigated axion-gravity systems coupled either to a massive dilaton or to a scalar field in a double well potential. In these models we found a whole zoo of wormhole solutions, both generalisations of Giddings-Strominger solutions and what we termed expanding wormholes. The former are characterised by the throat of the wormhole being a local minimum of the geometry, while the latter have a local maximum at the throat and a minimum  (or even several minima) further along the geometry. The expanding wormholes thus have a wineglass shape, which implies that after analytic continuation to Lorentzian time, they lead to expanding (rather than contracting) baby universes. 

In the scalar field case, we found that the presence of a potential barrier was essential in obtaining wormhole solutions. It is interesting that the existence of a potential barrier was also essential in the recent study, by two of us,  of scalar lumps with horizons \cite{Lavrelashvili:2021rxw,Lavrelashvili:2021qtg}, which are solutions that are relevant to gravitational tunnelling phenomena, in particular black hole seeded vacuum decay \cite{Hiscock:1987hn,Gregory:2013hja}. It would be important to see if suitable potential barriers can be obtained in string theoretic models. And, as for the scalar lumps, it would be worthwhile extending the wormhole solutions presented here to different asymptotics, in particular to quasi-de Sitter asymptotics.

The physical meaning of these classical wormhole solutions lies in the role they play in quantum gravity, namely
as saddle points of functional integrals. It is clear that in order for a classical solution to be valid, its associated saddle point must have a large action, $S_\text{E} \gg \hbar$. In the scalar field case, we found solutions with large action, but we also saw that the action can become small and even turn negative. Here, we may speculate that once the Euclidean action turns negative, the solution may no longer represent a relevant saddle point. This issue will however depend on the contours of integration of the gravitational path integral \cite{Feldbrugge:2017kzv}. It is a general property that a saddle point can only approximate a path integral if its weighting is smaller than the weighting of the geometries that are included in the definition of the integral. Hence, if the gravitational path integral is defined as summing over Lorentzian (pseudo-Riemannian) manifolds, which all have zero Euclidean action, then wormholes with negative Euclidean action (and thus larger weighting than Lorentzian geometries) will simply play no role \cite{Feldbrugge:2017mbc,Feldbrugge:2018gin}. However, we should point out that the definition of gravitational path integrals is still very much under debate (for a recent discussion see \cite{Lehners:2023yrj}).

Experience with metastable vacuum decay shows that there are three types of Euclidean solutions with finite action:
instantons \cite{Coleman:1978ae},
bounces \cite{Coleman:1977py}
and oscillating bounces \cite{Hackworth:2004xb}.
Instantons interpolate between vacua in a symmetric double well potential, have at most zero modes
and describe the mixing of these vacua.
Proper bounces in an asymmetric double well have a single negative mode in their spectrum of linear perturbations
\cite{Khvedelidze:2000cp,Koehn:2015hga},
and they describe metastable vacuum decay \cite{Coleman:1980aw}.
Oscillating bounces interpolate between vacua several times,
have multiple negative modes \cite{Lavrelashvili:2006cv,Battarra:2012vu} and their role in vacuum decay remains somewhat obscure.
The important question is to which class Euclidean wormholes belong and how we should interpret them.
There are several claims in the literature:
it was shown in \cite{Rubakov:1996cn} (see also \cite{Kim:1997dm,Kim:2003js}) that the GS wormhole has a single negative mode
in the lowest (homogeneous) sector. This statement was disproved in \cite{Alonso:2017avz}, where it was shown that with the proper choice of variables there
are no homogeneous negative modes about GS wormholes.
Later it was claimed that axionic Euclidean wormholes have multiple negative modes with higher angular harmonics \cite{Hertog:2018kbz}.
And recently this claim was disputed \cite{Loges:2022nuw} and it was demonstrated that GS wormholes do not have negative modes at all. We should point out that all of these studies concerned GS wormholes, and that the stability analyses so far included only the axion-gravity system and its fluctuations, without extra fields. 

To make progress, it will be important to incorporate the additional scalar or dilaton field. Our findings suggest that the stability analysis will be more involved in those cases. In particular, in the massive dilaton theory we found several different classes of solutions: there are the wormholes continuously connected to GS wormholes for small $m^2 N$ and
solutions with a ``mass gap'' which exist only above some minimal $m^2 N$ value (and which also exist at large dilaton couplings $\beta > \beta_c$). The latter arise at a bifurcation point. We also found expanding wormhole solutions, which are disconnected from GS wormholes, and also feature bifurcation points. Typically at such bifurcating points the stability properties of solutions change -- one branch usually picks up an additional negative mode \cite{Battarra:2013rba}. Let us recall here that negative modes signal that the solution under consideration is not a true minimum of the action, and that the action can be lowered (and the solution made more relevant) by deforming it in the direction indicated by the negative mode. The puzzling aspect in our case is that we find that typically solutions that have a more complicated field evolution actually have a smaller Euclidean action. The other way around would have appeared more plausible, with negative modes being associated with the extra features. Here, at least naively, it appears that there should be extra negative modes associated with the simpler field evolutions which, if true, would mark a surprising contrast with the purely axionic case. Clearly, an explicit linear stability analysis will be necessary to clarify this question. This will hopefully also elucidate the meaning of the multi-oscillation wormholes we have found. What might be helpful in addition would be to connect the stability analysis with catastrophe theory, which is a natural mathematical framework for describing such bifurcating behaviour.

We leave these intriguing questions for future investigation.

\acknowledgments
C.J. and J.-L.L. gratefully acknowledge the support of the European Research Council in the form of the ERC Consolidator Grant CoG 772295 ``Qosmology''.
The work of G.L. is supported in part by the Shota Rustaveli National Science Foundation of Georgia
with Grant N FR-21-860 and by the "EU fellowships for Georgian researchers" mobility programme.

\appendix

\section{Embedding diagrams}\label{sec:Embedding diagrams}

In order to visualise wormholes it is convenient to use embedding diagrams \cite{Misner:1973prb}.
Choosing the $h=1$ gauge in our ansatz \eqref{eq:sphericalansatz}, and fixing two angular variables on a unit sphere $\dd\Omega$ properly,
we obtain a two dimensional space described by the wormhole metric:
\be\label{eq:metric2}
\dd s^2 = \dd\tau^2 + a(\tau)^2 \dd\chi^2 \kma
\ee
where $\chi \in [0,2\pi]$ is the angular variable remaining.
We want to embed this two dimensional surface
in a three dimensional flat Euclidean space with cylindrical coordinates $\{Z, R, \chi\}$:
\be\label{eq:metric3}
\dd s^2 = \dd Z^2 + \dd R^2 + R^2 \dd \chi^2 \kma
\ee
If the embedded surface is described parametrically by $\{ R(\tau), Z(\tau)\}$, then the metric \eqref{eq:metric3} takes the form:
\be\label{eq:metric4}
\dd s^2 = ({\dot Z}^2 + \dot{R}^2) \dd R^2 + R^2(\tau) \dd \chi^2 \pkt
\ee
Comparing \eqref{eq:metric4} with the wormhole metric \eqref{eq:metric2}, we conclude that
\be\label{eq:ZR}
{\dot Z}^2 + {\dot R}^2=1 \kma~\quad R(\tau)= a(\tau) \pkt
\ee
Thus we can express $Z$ as
\be
Z(\tau)=\int_0^\tau d\tilde{\tau} \sqrt{1-{\dot a}^2(\tilde{\tau})}  \pkt
\ee
Then a revolution plot $\{a(\tau),Z(\tau)\}$ gives the wormhole visualisation as shown in Fig.~\ref{fig:embeddingdiagram}.

\section{Scaling of parameters} \label{sec:scaling}

Let us count how many free parameters we have.
\begin{enumerate}
		\item Massive dilaton: we may specify the wormhole charge to coupling ratio $q/f$ (or $N$), the dilaton mass $m$ and the dilatonic coupling constant $\beta$.
		\item Scalar field with double well potential: the action is specified by the wormhole charge to coupling ratio $q/f$ (or $N$) and the two potential parameters $\lambda$ and $v$.
\end{enumerate}
Under the following rescaling of the fields
\be
\phi \to \frac{\phi}{\sqrt{\kappa}}\,,\ a \to \frac{a}{\mu}\,,\ h \to \frac{h}{\mu}
\ee
and of the coupling constants to dimensionless variables
\be
q\to \frac{q}{\mu^2 \sqrt{\kappa}}\,,\  m\to \mu m\,,\ v \to \frac{v}{\sqrt{\kappa}}\,,\
\lambda \to \mu^2 \kappa \lambda\,,\
\ee
the dependance on $\mu$ and $\kappa$ is removed from the equations of motion and
the action scales as
\be
S \to \frac{1}{\kappa \mu^2} S = \frac{1}{\mu^2/M_\text{Pl}^2} S \kma
\ee
where $\mu$ is an arbitrary mass scale. In practice, while searching for solutions
this scaling freedom allows us to fix one parameter
({\it e.g.} $m$ or $N$ in the first theory and $\lambda$ or $N$ in the second theory) and vary the rest.

\section{Numerical estimation of wormhole action}\label{appendix:action_correction}

For numerical calculations we can divide the wormhole action into two pieces
\be
S_E = 2 \pi^2 \int_0^{\tau_*} \dd\tau  \left(\frac{2 N^2 \e^{-\beta \phi\sqrt{\kappa}}}{ a^3}  - a^3 V \right)
+ 2 \pi^2 \int_{\tau_*}^{\infty} \dd\tau  \left(\frac{2 N^2 \e^{-\beta \phi\sqrt{\kappa}}}{ a^3}  - a^3 V \right)
\equiv  S_E^{num} + S_E^{corr} \kma \label{eq:action_num}
\ee
where the first part is calculated numerically from $\tau =0$ to some big $\tau=\tau_*$
and the second part can be analytically estimated as follows.
Since we are interested in asymptotically flat solutions, $a(\tau) \to \tau$ as $\tau \to \infty$, the scalar field should go asymptotically to its ``vacuum'' value: $\phi = -v$ in the scalar field case and $\phi =0$ in the dilaton case.
In this setup the leading behaviour of the solution of the scalar equation Eq.~(\ref{eq:fulleomh1}) can be obtained analytically  and has the form:
\be
\phi(\tau) =  \frac{\beta N^2 \sqrt{\kappa}}{M^2 \tau^6} + C_1 \frac{\e^{-M \tau}}{\tau^{3/2}} + C_2 \frac{\e^{+M \tau}}{\tau^{3/2}} + ... \kma
\ee
where the dots denote corrections of higher order in $1/\tau,$ while $M$ is $\sqrt{2\lambda}v$ in the case of the scalar field and is $m$ in the case of the massive dilaton.
It is clear that for a regular solution the last, exponentially growing term should be absent.
This can be achieved by adjusting the initial value of the scalar field, while $C_2 \propto (\bar{\phi_0}-\phi_0)$,
where $\bar{\phi_0}$ is the initial value of the  scalar field of the actual regular solution.
Using this fact we can estimate $S_E^{corr}$ as
\be
S_E^{corr} = \frac{2\pi^2 N^2}{\tau_*^2}+ ... \kma
\ee
where the dots now denote corrections that are exponentially small and proportional to $\beta/\tau_*^8.$


\bibliographystyle{utphys}
\bibliography{biblio}

\end{document}